\documentclass[runningheads,a4paper]{llncs}

\usepackage{amssymb, amsmath}
\usepackage{graphicx}

\usepackage{url}

\usepackage{mathrsfs}
\usepackage{placeins}
\usepackage{subcaption}
\usepackage{graphicx}% Include figure files
\usepackage{dcolumn}% Align table columns on decimal point
\usepackage{bm}% bold math

\def\twoch{\,\,\begin{picture}(0,1) % ``2-chain''
\thicklines
\multiput(0,0)(0,10){2}{\circle*{2}}
\put(0,0){\line(0,1){10}}
\end{picture}\,\,}

\def\threech{\,\,\begin{picture}(0,1) % ``3-chain''
\thicklines
\multiput(0,0)(0,10){3}{\circle*{2}}
\put(0,0){\line(0,1){10}}
\put(0,10){\line(0,1){10}}
\end{picture}\,\,}

\def\twoach{\,\begin{picture}(1,1) % ``2-antichain''
\thicklines
\multiput(0,0)(5,0){2}{\circle*{2}}
\end{picture}\,\,\,}

\def\oneach{\,\begin{picture}(1,1) % ``1-antichain''
\thicklines
\multiput(0,0)(10,0){1}{\circle*{2}}
\end{picture}\,\,\,}

\def\Lcauset{\,\,\begin{picture}(2,2) % Lcauset ``L''
\thicklines
\multiput(0,0)(0,10){2}{\circle*{2}}
\put(0,0){\line(0,1){10}}
\multiput(0,0)(10,0){2}{\circle*{2}}
\end{picture}\,\,}

\def\lambdacauset{\,\begin{picture}(1,1) % ``lambda''
\thicklines
\multiput(0,0)(6.5,0){2}{\circle*{2}}
\multiput(3.25,10.2)(3.25,10.2){1}{\circle*{2}}
\put(0,0){\line(1,3){3.5}}
\put(6.5,0){\line(-1,3){3.5}}
\end{picture}\,\,\,}

\def\lambdafour{\,\begin{picture}(1,1) % ``4lambda''
\thicklines
\multiput(0,0)(6.5,0){2}{\circle*{2}}
\multiput(3.25,10.2)(3.25,10.2){1}{\circle*{2}}
\multiput(3.25,20)(3.25,20){1}{\circle*{2}}
\put(0,0){\line(1,3){3.5}}
\put(6.5,0){\line(-1,3){3.5}}
\put(3.25,10.2){\line(0,1){10}}
\end{picture}\,\,\,}

\def\fourch{\,\,\begin{picture}(0,1) % ``4-chain''
\thicklines
\multiput(0,0)(0,10){4}{\circle*{2}}
\put(0,0){\line(0,1){10}}
\put(0,10){\line(0,1){10}}
\put(0,20){\line(0,1){10}}
\end{picture}\,\,}

\def\vee{\,\,\begin{picture}(0,1) % ``V''
\thicklines
\put(0,0){\line(1,2){5}}
\put(0,0){\line(-1,2){5}}
\put(0,0){\circle*{2}}
\put(5,10){\circle*{2}}
\put(-5,10){\circle*{2}}
\end{picture}\,\,}

\def\diamond{\,\,\begin{picture}(0,1) % ``diamond''
\thicklines
\put(0,0){\line(1,2){5}}
\put(0,0){\line(-1,2){5}}
\put(0,0){\circle*{2}}
\put(0,20){\circle*{2}}
\put(5,10){\circle*{2}}
\put(-5,10){\circle*{2}}
\put(-5,10){\line(1,2){5}}
\put(5,10){\line(-1,2){5}}
\end{picture}\,\,}

\def\topvee{\,\,\begin{picture}(0,1) % ``top-V''
\thicklines
\put(0,10){\line(1,2){5}}
\put(0,10){\line(-1,2){5}}
\put(0,10){\circle*{2}}
\put(5,20){\circle*{2}}
\put(-5,20){\circle*{2}}
\put(0,0){\circle*{2}}
\put(0,0){\line(0,1){10}}
\end{picture}\,\,}

\newcommand{\lc}[1]{\tilde{#1}}
\newcommand{\Z}{\mathbb{Z}}
\newcommand{\N}{\mathbb{N}}

\begin{document}

\mainmatter  
\title{Covariant Growth Dynamics}

\author{Stav Zalel}

\institute{Blackett Laboratory, Imperial College, London, SW7 2AZ, U.K.}

\toctitle{Covariant Growth Dynamics}
\tocauthor{Authors' Instructions}
\maketitle

\begin{abstract} The spacetime discreteness of causal set theory has enabled the formulation of novel spacetime dynamics. In these so-called ``growth'' dynamics, a causal set spacetime is generated probabilistically by means of a random walk on certain tree structures. The first growth dynamics---the Classical Sequential Growth models---were proposed more than two decades ago and their study has furthered our understanding of general covariance and covariant observables within causal set theory. In this setting, labels take the place of spacetime coordinates so that general covariance takes the form of label-invariance and covariant observables are those order-theoretic properties of the causal set which are label-independent. In recent years, these insights have led to a new formulation of growth dynamics which makes no reference to labels. Here we present an overview of these (manifestly) covariant growth dynamics.

\end{abstract}

\section{Introduction}
In causal set theory---where the continuum spacetime of General Relativity is replaced by a discrete causal set---spacetime points and their coordinates are replaced by causal set elements and their ``labels''. Thus, Einstein's struggle between the formulation of generally covariant laws of nature and the intrinsic (in)distinguishability of spacetime points \cite{Schilpp:1949,Stachel:2014,Rovelli:1990ph} manifests in causal set theory as tension between label-dependence and label-independence. This tension can be expressed via a myriad of interrelated questions: What is the physical status which one should assign to the causal set elements and to their labels? Should we conflate the ``label'' of an element with the ``intrinsic identity'' of an element or should they be considered separately? What are the precise mathematical concepts which are best suited for formulating a physical theory of causal sets? In particular, could the theory be formulated without any reference to ``labels''?

The importance of understanding general covariance in any given theory is heightened within the path integral approach to quantum dynamics \cite{PhysRevD.18.1747,Sorkin:1997gi,Gibbons:2006pa} whose most appealing feature in the context of gravity is arguably its compatibility with general covariance: the integral sums over \textit{complete} spacetime histories and therefore does not require a foliation or a distinguished time parameter, while covariant ``observables''  can be defined independently of observers as attributes of histories.

Defining the path integral for quantum gravity remains a challenge \cite{GIBBONS1987736,GIBBONS1978141}, and one may be justified in regarding the path integral as a guiding principle rather than an exact prescription. In causal set theory, the path integral is replaced by a discrete ``sum-over-histories''. The challenges in its definition and interpretation may be summarised by 3 open questions:
 \begin{itemize} \item[] \textit{What is the domain of the sum-over-histories?}\vspace{1mm}
\item[] \textit{What is the amplitude by which each history should be weighted?}\vspace{1mm}

\item[] \textit{What are the physical observables?}
\end{itemize}

These are the questions that the growth dynamics program aims to answer \cite{Rideout:1999ub,Brightwell:2011,Brightwell:2012,Brightwell:2002yu,Brightwell:2002vw,Varadarajan:2005gg,Dowker:2005gj,evidence,Georgiou:2010,Dowker:2010qh,Dowker:2014xga,Wuthrich:2015vva,Sorkin:2007hga,Dowker:2020qqs,Bombelli:2008kr,Sorkin:1998hi,Martin:2000js,Ash:2002un,Ash:2005za,Ahmed:2009qm,Dowker:2017zqj}. A growth dynamics is a probabilistic process in which a causal set comes into being \textit{ex nihilo} by accretion of elements. This growth process plays a dual role: it embodies the sum-over-histories (\textit{e.g.} by providing  a mechanism from which the action is emergent) and it offers a novel route for accounting for the passage of time within physics \cite{Dowker:2014xga,Wuthrich:2015vva,Sorkin:2007hga,Dowker:2020qqs,Norton_2018}. Crucially, the growth does not happen \textit{in} time---it constitutes the passage of time. The birth of an element is the \textit{happening} of that event, while the existence of an element signifies that the event has \textit{already happened}. Thus heuristically, the growth process is a physical process whose phenomenological manifestations is the passage of time.\footnote{While work in the field is usually motivated by this interpretation, its results are not contingent on it.}

The archetype of growth dynamics for causal sets are the Classical Sequential Growth (CSG) models of \cite{Rideout:1999ub}. In these models, the causal set elements are born one after another, (in a sequence, hence the name \textit{sequential}) and form relations with each other according to model-dependent probabilities. The functional form of the probabilities satisfies mathematical constraints motivated by local causality and general covariance. The latter, known as \textit{discrete general covariance}, states that the probability that the first $n$ born elements form some causal set $\lc{C}_n$ is equal to the probability that they form any causal set which is order-isomorphic to $\lc{C}_n$.

In the language of growth dynamics, the \textit{domain} of the sum-over-histories is the sample space of the growth process (\textit{i.e.} the causal sets or \textit{histories} which can be grown by the process). The role of the \textit{amplitude} by which each history is weighted is played by the probabilities which govern the stochastic growth of the causal
set. And the \textit{observables} are sets of histories (known in the probability literature as \textit{events}). Thus, the open questions which are posed by the sum-over-histories take on a more concrete form: 
\begin{itemize}
\item[] \textit{Should the growth process produce ``labeled'' or ``unlabeled'' causal sets? Should it produce all infinite causal sets or only those which are past-finite?}\vspace{1mm}
\item[] \textit{How should the probabilities be constrained to obtain physical dynamics? And how can the probabilities be generalised into complex amplitudes so that the resulting dynamics exhibits quantum interference?}\vspace{1mm}
\item[] \textit{Which events or observables have a physical interpretation? Which of these should we interpret as ``local'' and which as ``global''?}
\end{itemize}

In this chapter we take our cue from the first of these questions, reviewing the state-of-the-art growth dynamics for ``unlabeled'' causal sets. We call these dynamics \textit{manifestly covariant dynamics}, or simply \textit{covariant dynamics} for short. But our discussion will touch upon all the above themes, revealing the interconnectedness of domain, amplitude and observables.

From the start, one can identify three challenges. First, unlabeled causal sets are mathematically more difficult to handle (\textit{e.g.} enumerating unlabeled graphs is generally more difficult than enumerating labeled graphs \cite{Harary:1973}), echoing the difficulties often encountered in physics when working with global degrees of freedom. Second, a common approach to obtaining a physical dynamics is to impose invariance under certain gauge transformations. Indeed, the CSG models were obtained by imposing the discrete general covariance condition which is akin to the requirement that the Einstein-Hilbert action is invariant under diffeomorphisms. Therefore even if a label-independent framework did exist at the level of the kinematics, how are the physically meaningful dynamics to be picked out from the plethora of available models? Finally, our intuitive notion of growth is inherently sequential: elements are born one after the other in a kind of global time which renders the elements distinguishable (\textit{e.g.} in a CSG model, each element is labeled by its position in the sequence of births). How does one reconcile a physical process of becoming with manifest covariance? The prevailing view in causal set theory is that one should seek a form of ``asynchronous becoming'', namely a growth process in which elements are born in a partial (not a total) order \cite{Dowker:2014xga,Sorkin:2007hga,Dowker:2020qqs}. What could it mean for elements to be born in a partial order? It is the role of mathematics to make sense of notions which lie beyond our everyday experience, and it may be that new mathematics is what is needed to better understand asynchronous becoming and its consequences for the nature of time. It has been suggested in \cite{Wuthrich:2015vva} that this could be achieved via a ``novel and exotic'' framework in which questions such as ``which element is born at stage $n$?'' are left unanswered (not because of ignorance, but because they are unphysical). As we shall see, this notion is affirmed by our current understanding of covariant dynamics \cite{Dowker:2019qiz,Zalel:2020oyf,Bento:2021voo}.

\section{Labels and label-invariance}\label{chapter_labels}

Should we conceive of the elements of a causal set as distinguishable or indistinguishable? Mathematically, the elements of a causal set are distinguishable (in so far as a causal set is a set) and the notion of labeling these elements appears in pure mathematics and in its applications, including in causal set theory where labels naturally arise within the CSG models. But the clear distinguishable-indistinguishable divide is blurred by opposing physical notions within causal set theory. On the one hand, the correspondence between spacetime volume and the number of spacetime elements forces one ``to accept that the elements of the causal set are real, and that volume measurements ``count'' them in much the same way that weighing a copper ingot ``counts" the number of atoms it comprises'' \cite{Sorkin:1989re}. On the other hand, it is a postulate of causal set theory that no information is contained in any individual identity or label of the elements, so that stripped of ordering the elements are physically indistinguishable. Is this indistinguishability compatible with the elements' physical existence? Can you measure the cardinality of an antichain? These subtle questions underpin the discussion of covarinace and labels within causal set theory. Largely, the answers given by the community to these last two questions have been a yes. Considered separately, each element has no distinguishing characteristic, but arranged together in a partial order they give rise to a meaningful structure whose properties include, for example, a notion of cardinality. These considerations, as well as the analogy between coordinates in the continuum and labels in the discrete, have led to the understanding of general covariance as label-invariance within causal set theory \cite{Rideout:1999ub,Brightwell:2002yu,Brightwell:2002vw}. This understanding is our starting point for the discussion of covariant dynamics.

\subsection{The notion of labeling}\label{sec_hist_lab}

A \textit{partial order} is a pair, $(\Pi, \prec)$, where $\prec$ is a transitive, irreflexive relation on the ground-set $\Pi$. A \textit{linear order} (also known as a \textit{total order}) is a partial order in which any two elements are comparable. 
A \textit{labeling} of a set $\Pi$ is a mapping $\lambda$ from $\Pi$ to an index set $\mathcal{I}$. When $\Pi$ carries additional structure, it may be desirable that the labeling reflect this additional structure. In particular, when labeling a partially ordered set $(\Pi,\prec)$ one often endows the index set with a total order $\ll$ and requires that the labeling is order-preserving, \textit{i.e.} $x\prec y \implies \lambda(x)\ll \lambda(y) \ \forall x,y\in\Pi$. Thus, a labeling of $(\Pi,\prec)$ arranges the elements of $\Pi$ into a linear order $\ll$ which is compatible with the partial order $\prec$. This notion of labeling is shared by the various definitions that can be found in the literature \cite{BRIGHTWELL199487,Alon:1994,BRIGHTWELL1988113}.

An important special case is the \textit{natural labeling} where the index set of labels is (a subset of) the natural numbers. In words, a natural labeling is an enumeration of the causal set elements which respects the partial order. Thus while the labels contain an element of gauge, they reflect the partial order through their compatibility with it. We may draw an analogy with a familiar example from the continuum: inertial coordinates on Minkowski spacetime. Inertial coordinates provide a labeling of spacetime points, where each spacetime point $p$ is labeled by a coordinate vector $(t_p, \bar{x}_p)$. The coordinates reflect the causal structure through the time coordinate, since if a spacetime point $p$ is in the causal past of another spacetime point $q$ then $t_p<t_q$.

Equipped with the notion of labeling, one may use the term \textit{labeled partial order} to mean a partial order together with a natural labeling of it \cite{Brightwell:2002vw}. In practice, one often discusses labeled partial orders without specifying the ground-set $\Pi$ by repackaging the information contained in the order relation $\prec$ and the natural labeling into a partial order on a set of natural numbers \cite{Brightwell:2002vw,Ash:2002un,Sorkin:2011sp}.

The term \textit{unlabeled partial order} is borrowed from graph theory and means that the elements of the partial order are indistinguishable when stripped of the order relations. Therefore, an unlabeled partial order is not in fact a partial order but an order-isomorphism equivalence class of partial orders. Which partial orders are contained in a given equivalence class depends on one's universe of discourse (\textit{e.g.} partial orders on a specified ground-set).

\subsection{Terminology}\label{terminology_1}
A \textit{causal set} (or \textit{causet}) is a locally finite partial order. For any natural number $n<\infty$, let $[0,n]$ denote the set $\{0,1,...,n\}$ (devoid of any ordering).

\begin{definition}[Labeled causet]\label{def_labeled_causet}
A labeled causet is any causet $([0,n],\prec)$ or $(\mathbb{N},\prec)$ satisfying $x\prec y \implies x<y$.
\end{definition}

\begin{definition}[$n$-causet]
An $n$-causet is a labeled causet of cardinality $n$.
\end{definition}

Our universe of discourse contains all labeled causets and their subcausets. (Note that a subcauset of a labeled causet is not necessarily a labeled causet because its ground-set may not be an interval of integers of the form $\{0,1,...,n\}$.) Given some $n>0$, we denote the set of $n$-causets by $\tilde{\Omega}(n)$. The set of finite labeled causets is denoted by $\tilde{\Omega}(\mathbb{N})$, \textit{i.e.} $\tilde{\Omega}(\mathbb{N}):=\bigcup \limits_{n>0}\tilde{\Omega}(n)$.The set of infinite labeled causets is denoted by $\tilde{\Omega}$. Labeled causets and their subcausets are denoted by capital Roman letters with a tilde, \textit{e.g.} $\lc{C}$. We often (but not always) use a subscript to denote the cardinality of an $n$-causet, \textit{e.g.} $\lc{C}_n$.

\begin{definition}[Order] An order (or ``unlabeled causet'') is an order-isomorphism equivalence class of labeled causets.\end{definition}

\begin{figure}[htpb]
  \centering
	\includegraphics[width=0.6\textwidth]{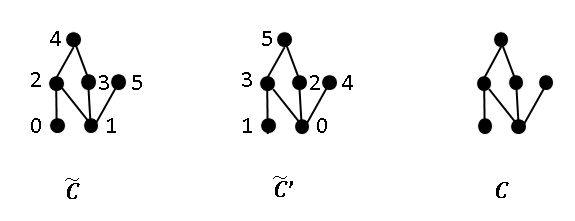}
	\caption[Labeled causets and orders]{$\lc{C}$ and $\lc{C}'$ are order-isomorphic labeled causets. Each is a representative of the order $C$, shown on the right as a Hasse diagram without labels.}
	\label{order_example} 
\end{figure}

We denote orders by capital Roman letters without a tilde. Given an order, the Hasse diagrams of its representatives differ from each other only by the labeling of nodes (\textit{i.e.} they are graph-isomorphic). Therefore, we represent an order by a Hasse diagram without node labels (Fig.\ref{order_example}).

\begin{definition}[Cardinality of an order] The cardinality of an order is defined to be the cardinality of a representative of it. \end{definition}

We denote the cardinality of an order $C$ by $|C|$.

\begin{definition}[$n$-order] An $n$-order is an order of cardinality $n$. \end{definition}
In other words, an $n$-order is an order whose representatives are $n$-causets. We often (but not always) use a subscript to specify the cardinality of an $n$-order, \textit{e.g.} $C_n$. 

We use $\Omega(n)$, ${\Omega}(\mathbb{N})$ and $\Omega$ to denote the set of $n$-orders, the set of finite orders and the set of infinite orders, respectively. Note that these are equivalent to the quotient spaces $\lc{\Omega}(n)/\cong$, $\lc{\Omega}(\mathbb{N})/\cong$ and $\lc{\Omega}/\cong$, where $\cong$ denotes equivalence under order-isomorphism.

Similarly to the way an order ``inherits'' the cardinality of its representatives, an order is future-finite if its representatives are future-finite \textit{etc.} We may also refer to an element of an order, meaning an element of a representative of it---the meaning should be clear from the context.

\begin{definition}[Stem---\textit{labeled}] A stem in a labeled causet $\lc{C}$ is a finite subcauset $\lc{D}\subseteq\lc{C}$ which contains its own past, \textit{i.e.} if $x\in\lc{D}$ and $y\prec x$ in $\lc{C}$ then $y\in\lc{D}$. \end{definition}
In particular, given any labeled causet $\lc{C}$ and an integer $n$ satisfying $0\leq n\leq |\lc{C}|$, the restriction of $\lc{C}$ to the interval $[0,n]$ is a stem in $\lc{C}$.

\begin{figure}[htpb]
  \centering
	\includegraphics[width=0.8\textwidth]{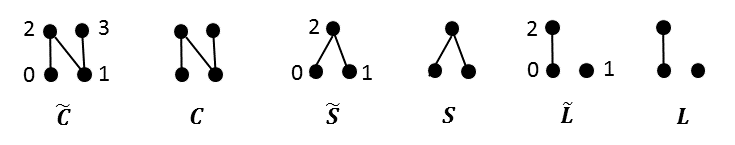}
	\caption[Stems]{Labeled causets  $\lc{C}, \ \lc{S}$ and $\lc{L}$ are representatives of orders $C,\ S$ and $L$, respectively. $\lc{S}$ is a stem in $\lc{C}$. $\lc{L}$ is not a subcauset of $\lc{C}$ so it is not a stem in $\lc{C}$. $S$ and $L$ are stems in $\lc{C}$ and in $C$. }
	\label{defn_unlabeled_stem} 
\end{figure}

\begin{definition}[Stem---\textit{unlabeled}] A finite order $S$ is a stem in the order $C$ if there exists a representative of $S$ which is a stem in some representative of $C$. When $S$ is a stem in $C$, we may also say that $S$ is a stem in any representative $\lc{C}$ of $C$. \end{definition}
Hence, the meaning of ``stem'' depends on the context (Fig.\ref{defn_unlabeled_stem}).

\begin{definition}[$n$-stem] An $n$-stem is a stem of cardinality $n$.\end{definition}
 
The notion of \textit{rogue} is closely related to that of stem.

\begin{definition}[Rogue---\textit{labeled}] An infinite labeled causet $\lc{C}\in\lc{\Omega}$ is a rogue if there exists some $\lc{D}\in\lc{\Omega}$ such that $\lc{C}\not\cong\lc{D}$ and $S\in\Omega(n)$ is a stem in $\lc{D}$ if and only if $S$ is a stem in $\lc{C}$. We say that $\lc{C}$ and $\lc{D}$ are \textit{equivalent rogues} or a \textit{rogue pair}. \end{definition}

\begin{figure}[htpb]
  \centering
	\includegraphics[width=0.5\textwidth]{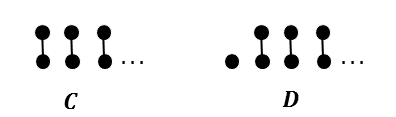}
	\caption[Rogue pair]{$C$ is a countable union of 2-chains and $D$ is the union of $C$ with a single unrelated element.  $C$ and $D$ have the same stems---any union of finitely many 2-chains and a finite, unrelated antichain---hence, $C$ and $D$ are equivalent rogues.}
	\label{rogueexample}
\end{figure}

\begin{definition}[Rogue---\textit{unlabeled}]\label{rog_def} An order is a rogue if its representatives are rogues. Equivalently, $C$ and $D$ are a rogue pair when $S\in\Omega(n)$ is a stem in $D$ if and only if $S$ is a stem in $C$. \end{definition}
Rogue equivalence, denoted by $C\sim_R D$, is an equivalence relation on $\Omega$. An example is shown in Fig.\ref{rogueexample}.

\subsection{Growth dynamics}\label{sec_gd}
\begin{figure}[htpb]
 \centering
	\includegraphics[width=0.7\textwidth]{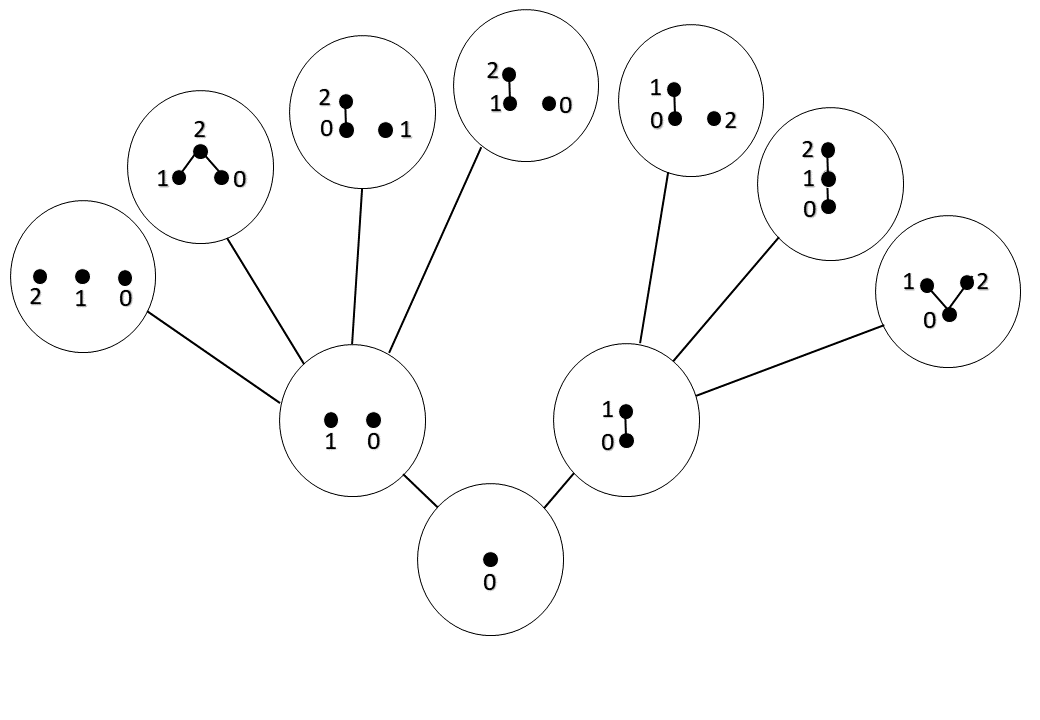}
	\caption{Sequential growth. Elements are labeled by the stage at which they are born.}
	\label{lposcau}
\end{figure}

What role do labels play within the growth dynamics framework? In the sequential growth models, elements are born one by one, and so they are labeled by the stage at which they are born (Fig.\ref{lposcau}). Thus, each realisation is a labeled causet and the sample space (allowing the process to continue \textit{ad infinitum}) is the set of infinite labeled causets, $\lc{\Omega}$. But our understanding of labels as pure gauge suggests that only the covariant statements that we can make about these realisations are physical. We now make this notion precise.

For each finite labeled causet $\lc{C}_n$, define the ``cylinder set'',
\begin{equation}\label{cyl_def_chap1}cyl(\lc{C}_n):=\{\lc{C}\in\tilde{\Omega}|\lc{C}_n \text{ is a stem in } \lc{C}\}.\end{equation}
We denote the $\sigma$-algebra generated by the cylinder sets by $\lc{\mathcal{R}}$. $(\lc{\Omega},\lc{\mathcal{R}})$ is a measurable space on which each sequential growth model induces a unique probability measure $\lc{\mu}$ satisfying,
\begin{equation}\label{eq_premeasure_sec_1}
\lc{\mu}(cyl(\lc{C}_n))=\mathbb{P}(\lc{C}_n) \  \forall \ \lc{C}_n\in\lc{\Omega}(\mathbb{N}),
\end{equation}
where $\mathbb{P}(\lc{C}_n)$ is the model-dependent probability that the first $n$ elements form $\lc{C}_n$. The existence of $\lc{\mu}$ is guaranteed by the so-called \textit{extension theorem} of measure theory which states that a pre-measure on a semiring (the cylinder sets form a semiring on which \eqref{eq_premeasure_sec_1} defines a pre-measure) extends to a measure on the $\sigma$-algebra generated by the semiring \cite{Kolmogorov:1975}.

Using the terminology of measure theory, each set of histories $\lc{\mathcal{E}}\in\lc{\mathcal{R}}$ is an \textit{event}. Physically, the interpretation of these events are as \textit{observables} or \textit{beables}. Thus, we are only interested in the covariant events, defined as follows.

\begin{definition}[Covariant event] An event $\lc{\mathcal{E}}\in\lc{\mathcal{R}}$ is covariant if whenever $\lc{C}\in\lc{\mathcal{E}}$ and $\lc{C}\cong\lc{D}$ then $\lc{D}\in\lc{\mathcal{E}}$. \end{definition}

In words, a covariant event is one which cannot distinguish between order-isomorphic causets. 
The collection of covariant events is a $\sigma$-algebra (a sub-$\sigma$-algebra of $\tilde{\mathcal{R}}$) and we denote it by $\mathcal{R}$.

One can conceive of $\mathcal{R}$ as a $\sigma$-algebra on $\Omega$, the set of infinite orders, via the projection $p:\lc{\Omega}\rightarrow\Omega$ which assigns to each causet $\lc{C}$ the order $C$ of which it is a representative. Whether a covariant event is a set of causets or a set of orders is of no consequence for our purposes and we will use the two interchangeably depending on which is more convinient in the context.

In light of this, one may formulate the problem of covariant dynamics as a two-part question: Can a measure be defined directly on $\mathcal{R}$ (without using $\tilde{\mathcal{R}}$ as an intermediary)? And if so, can it be done by means of a random walk whose sense of dynamical progression one may interpret as growth? Since one usually considers random walks on finite valency trees which give rise to a countable semiring of cylinder sets, one technical question which underpins this discussion is whether $\mathcal{R}$ is countably generated. As far as the author is aware, this is an open question.

Instead, the direction which has been pursued by the community has been to study sub-$\sigma$-algebras of $\mathcal{R}$ \cite{Brightwell:2002yu,Brightwell:2002vw,Dowker:2005gj}. This approach has several advantages. First, the sub-$\sigma$-algebras of interest have been shown to be isomorphic to the topological $\sigma$-algebras of certain trees. This means that one can define a measure on them by means of a random walk on a tree, allowing for a growth dynamics picture. Second, while all the events in $\mathcal{R}$ are covariant, the physical interpretation of these events remains largely obscure. But in some sub-$\sigma$-algebras all events can be assigned a physical interpretation, an argument for considering them alone as the physical set of observables. Third, this kinematic argument for narrowing the set of observables is strengthened by a dynamical one. Consider some measure $\mu$ on $\mathcal{R}$ which satisfies $\mu(\mathcal{E})=0$ for some $\mathcal{E}\in\mathcal{R}$. Then the measure of an arbitrary event $\mathcal{F}\in\mathcal{R}$ is fixed via $\mu(\mathcal{F})=\mu(\mathcal{F}\setminus\mathcal{E})$. The collection of events of the form $\mathcal{F}\setminus\mathcal{E}$ is contained in some sub-$\sigma$-algebra of $\mathcal{R}$ which we denote as $\mathcal{R}_{\mathcal{E}}\subset\mathcal{R}$. Physically, we can interpret this statement as saying that the events in $\mathcal{R}\setminus \mathcal{R}_{\mathcal{E}}$ contain no new dynamical information and therefore $\mathcal{R}_{\mathcal{E}}$ exhausts the set of observables for the particular dynamics $\mu$.

The CSG models make a good case-study for all three arguments. Proceeding in reverse order, in the CSG models the set of all rogues (the event that spacetime is a rogue) has measure zero and the measure on the \textit{stem algebra} (defined below) is sufficient to recover the measure on $\mathcal{R}$ \cite{Brightwell:2002vw}.

For each $n$-order $C_n$, its \textit{stem set} is defined as,
\begin{equation}\label{stem_set_def}\begin{split}
stem(C_n) := &\{  \lc{D} \in \tilde{\Omega} \ |\  C_n \ \textrm{is a stem in} \ \lc{D} \},
\end{split}
\end{equation}
and is equal to the union of cylinder sets of the labeled causets in which $C_n$ is a stem.
We denote the $\sigma$-algebra generated by the stem sets by $\mathcal{R}(\mathcal{S})$ and note that $\mathcal{R}(\mathcal{S})\subset\mathcal{R}$.

In fact, one can identify strictly smaller $\sigma$-algebras in $\mathcal{R}(\mathcal{S})$ from which the measure on $\mathcal{R}$ can be recovered. But there is a strong kinematic argument for crowning $\mathcal{R}(\mathcal{S})$ as the physical set of observables: each event in $\mathcal{R}(\mathcal{S})$ has a physically meaningful interpretation as a logical combination of statements about which finite orders are stems in the growing causal set (\textit{e.g.} the event $stem(\twoach)\cap stem(\twoch)$ corresonds to the statement \twoch and \twoach \ \ are both stems in the growing causet). Therefore, one can characterise the events in $\mathcal{R}(\mathcal{S})$ as those covariant events which do not distinguish between equivalent rogues.

\begin{figure}[htpb]
 \centering
	\includegraphics[width=0.63\textwidth]{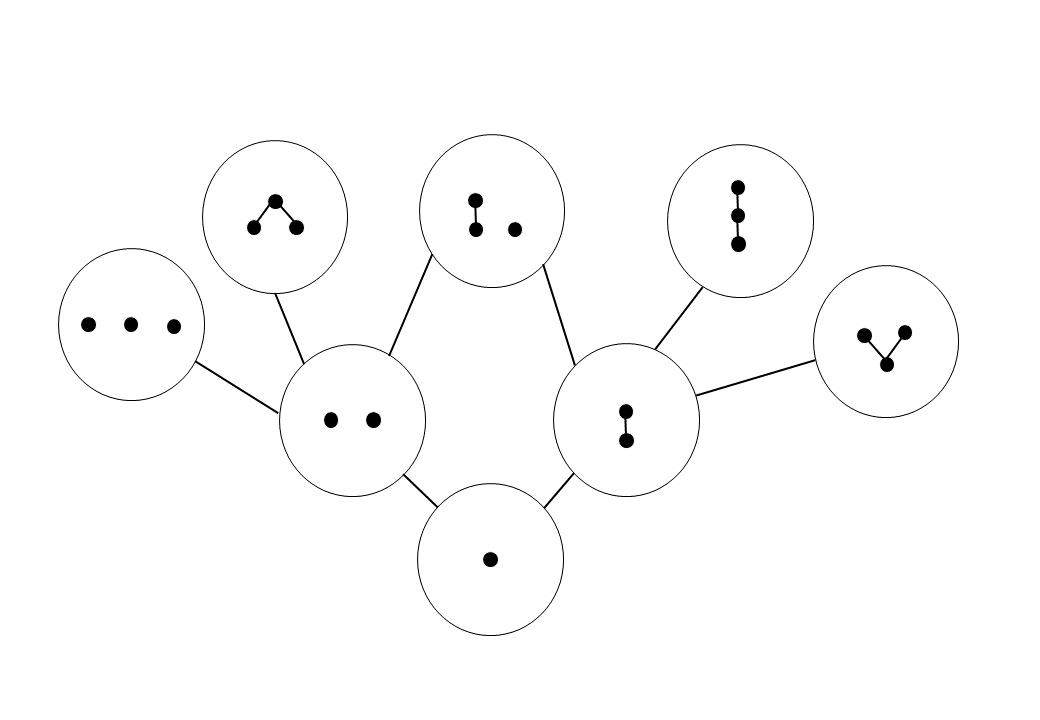}
	\caption[Poscau]{The first three levels of poscau.}
	\label{hasse}
\end{figure}

Finally, can we conceive of a measure on $\mathcal{R}(\mathcal{S})$ in terms of a random walk? The stem sets can be arranged into a partial order by means of set inclusion (\textit{i.e} $stem(C)\prec stem(D)$ if $stem(C)\supset stem(D)$). This ordering of the stem sets is equivalent to \textit{poscau} (Fig.\ref{hasse}).
\begin{definition}[Poscau] Poscau is a partial order on finite orders, $(\Omega(\mathbb{N}), \prec)$, where $A\prec B$ if and only if $A$ is a stem in $B$.\end{definition}
To conceive of a random walk on poscau as a physical process, each node should carry a clear physical meaning. Naively, arriving at the node $A$ corresponds to the occurrence of the covariant event $stem(A)$. But this fails because the physical interpretation of the nodes implies that each growing causal set contains only one $n$-stem for each $n>0$ (which is untrue in general). This failure is routed in the fact that poscau is not a tree which is closely related to the fact that the collection of stem sets does not behave like a collection of cylinder sets, since $stem(A)\cap stem(B)\not= \emptyset$ for all $A,B$.
 
\section{Introducing covtree}\label{chapter_covtree_4}

We discussed the difficulty of assigning physical meaning to a walk on poscau, and thinking in this way suggests the solution: a covariant dynamics can be defined as a walk on a tree formed of countably many levels in which the nodes in level $n$ are not single $n$-orders but \textit{sets} of $n$-orders. Each set of  $n$-orders in level $n$ will correspond to the covariant event ``the $n$-stems of the growing causal set are the elements of this set.''  We call this tree \textit{covtree}, short for \textit{covariant tree}.

All definitions and results presented in this section are taken from \cite{Dowker:2019qiz}.

\subsection{Certificates}\label{cert_Sec}

We now introduce the notion of \textit{certificate} which will play a key role in the definition of covtree and in its interpretation as a framework for growth dynamics.

\vspace{2mm}
Let $\Gamma_n\subseteq \Omega(n)$ be a non-empty set of $n$-orders. 
\begin{definition}[Certificate]\label{def_cert_sec_2}
A finite or infinite order $C$ is a certificate of $\Gamma_n$ if $\Gamma_n$ is the set of  all $n$-stems in $C$.
\end{definition}

Given some $\Gamma_n$, it may or may not have a certificate. We will be interested in those $\Gamma_n$ which do have a certificate.

\begin{definition}[$\Lambda$, the collection of certified sets]\label{def_cert_sec_2}
$\Lambda$ is the collection of sets of $n$-orders, for all $n$, for which there exists a certificate:
\begin{equation}\label{lambda_def_chap2}
\Lambda := \bigcup\limits_{n>0}\{\Gamma_n \subseteq \Omega(n)| \exists \text{ a certificate for } \Gamma_n\}. 
\end{equation}
\end{definition}

One can show that each $\Gamma_n\in\Lambda$ has infinitely many certificates, including infinitely many finite certificates and infinitely many infinite certificates. We will often work with the \textit{minimal} certificates:

\begin{definition}[Minimal certificate]\label{min_cert_def} Given some $\Gamma_n\in\Lambda$, we order its finite certificates as follows: let $C,C'$ be finite certificates of $\Gamma_n$, then $C\preceq C'$ if $C$ is a stem in $C'$. A minimal certificate of $\Gamma_n$ is minimal in this partial order of certificates.
\end{definition}

At times it may be easier to work with labeled causets rather than with orders. To this end we define the labeled analogue of the certificate.
\begin{definition}[Labeled certificate]
A labeled certificate of $\Gamma_n$ is a representative of a certificate of $\Gamma_n$. A labeled minimal certificate of $\Gamma_n$ is a representative of a minimal certificate of $\Gamma_n$.
\end{definition}

\begin{figure}[htpb]
  \centering
	\includegraphics[width=0.7\textwidth]{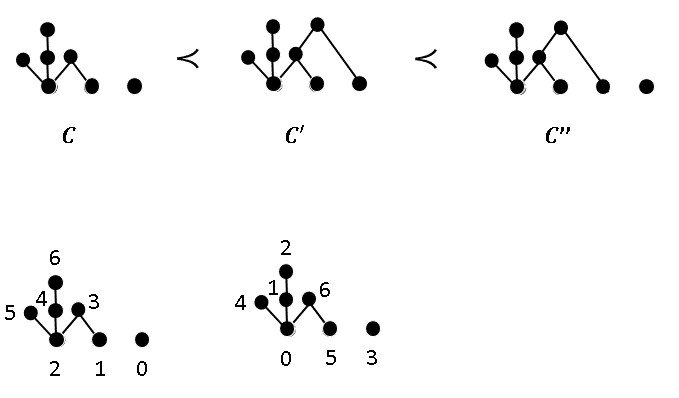}
	\caption[Certificates]{Certificates. $C$, $C'$ and $C''$ are finite certificates of  $\Omega(3)$, the set of all 3-orders. The $\prec$ relation indicates inclusion by stem. $C$ is a minimal certificate of $\Omega(3)$. The labeled causets shown are representatives of $C$ and hence are labeled minimal certificates of $\Omega(3)$.}
	\label{certificate_figure}
\end{figure}

Illustrations are shown in Fig.\ref{certificate_figure}.

\FloatBarrier

\subsection{Definition of covtree}\label{subseccovtree}

We begin by introducing the map $\mathcal{O}$.

\begin{definition}[The map $\mathcal{O}$]\label{o_minus_defn} For any $n>1$ and any $\Gamma_n$, the map $\mathcal{O}$  takes $\Gamma_n$ to the set of $(n-1)$-stems of elements of $\Gamma_n$:
\begin{equation} 
\mathcal{O}(\Gamma_n):=\{B_{n-1} \in \Omega(n-1)\ | \  \exists \ A_n\in \Gamma_n\ \mathrm{ s.t. }\ B_{n-1} \text{ is a stem in } A_n \}
\,. 
\end{equation} 
\end{definition}

\begin{figure}[htbp]
    \centering
    \includegraphics[width=.5\textwidth]{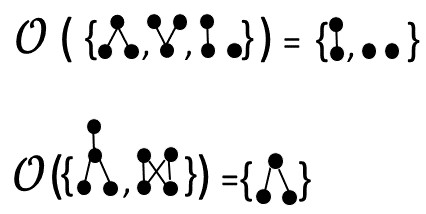}
    \caption[The map $\mathcal{O}$]{Illustration of the map $\mathcal{O}$.}
    \label{o_definition}
\end{figure}

An illustration is shown in Fig.\ref{o_definition}. The exponentiation $\mathcal{O}^k$ takes $\Gamma_n$ to the set of $(n-k)$-stems of elements of $\Gamma_n$. If $C$ is a certificate of $\Gamma_{n}$, then $C$ is also a certificate of ${{\mathcal O}}^k (\Gamma_{n}) $ for any $k<n$.\footnote{The proof may be summarised by the mnemonic: a stem in a stem is a stem, not a stem in any stem is not a stem} The converse is not true: if $C$ is a certificate of ${\mathcal O}(\Gamma_{n})$, then $C$ may or may not be a certificate of $\Gamma_{n}$ (in fact, $\Gamma_{n}$ may have no certificates at all).

\begin{definition}[Covtree]\label{covtree_def} Covtree is the partial order $(\Lambda, \prec)$, where $\Gamma_n\prec\Gamma_{m}$ if and only if $n<m$ and ${\mathcal{O}}^{m-n}(\Gamma_m)=\Gamma_n$. \end{definition}

\begin{figure}[htbp]
    \centering
    \begin{subfigure}{10cm}
    \includegraphics[width=.9\textwidth]{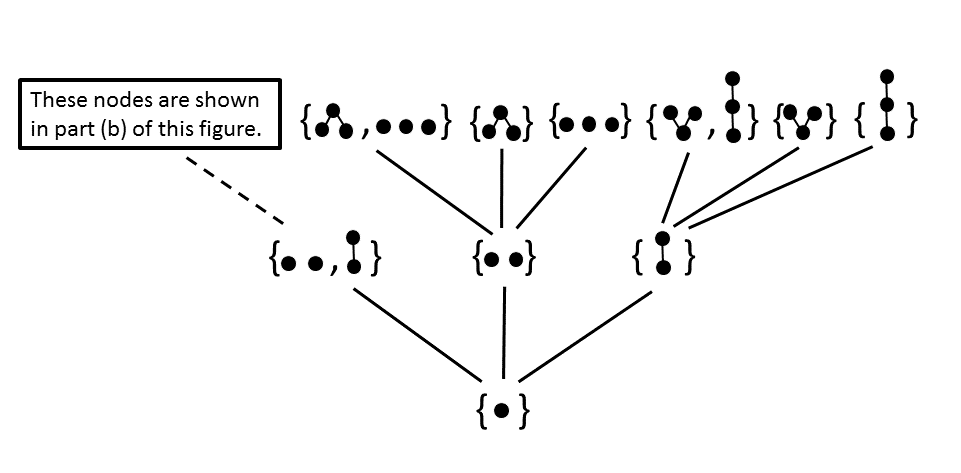}
    \caption{The structure of the first three levels of covtree.}
    \end{subfigure}

    \begin{subfigure}{10cm}
   \includegraphics[width=0.9\textwidth]{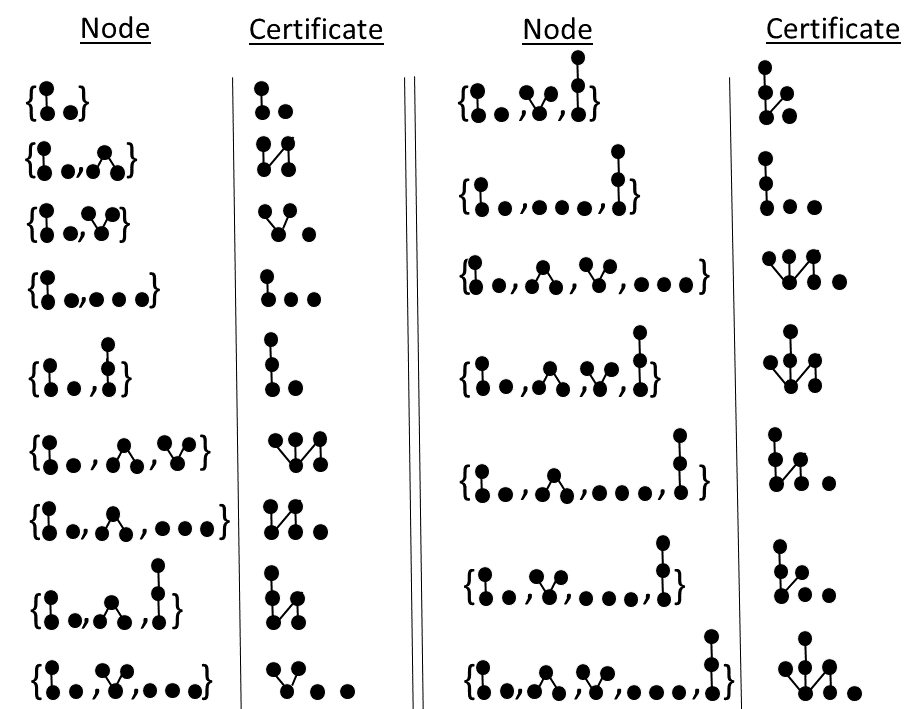}
	\caption{The level 3 nodes which are directly above the node $\{$\twoch, \twoach \ $\}$ are shown together with their respective certificates.}
    \end{subfigure}
    \caption[Covtree]{The first three levels of covtree.}%
    \label{covtree_fig_chap_2}%
\end{figure}

\begin{figure}[htpb]
  \centering
	\includegraphics[width=0.5\textwidth]{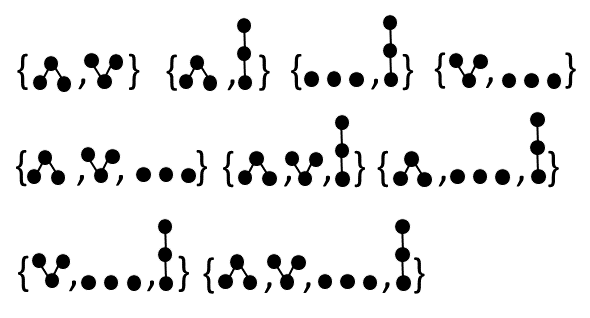}
	\caption[Non-nodes]{The sets shown in the figure have no certificates and therefore are not nodes. For every set shown, if an order contains all the elements of that set as stems then it also contains \hspace{1mm} \Lcauset \hspace{2mm} as a stem.}
	\label{nonnodes}
\end{figure}

We note some key points about covtree:
\begin{itemize}
\item Covtree is the partial order on $\Lambda$ defined by putting each $\Gamma_n$ directly above $\mathcal{O}(\Gamma_n)$ and taking the transitive closure. Thus, covtree is a tree.

\item Covtree has no maximal nodes. Every covtree node is contained in uncountably many inextendible upward-going paths.

\item We label the levels of covtree by 1,2,... where level 1 contains the root. The nodes at level $n$ are the sets of $n$-orders which have certificates (this is the motivation for the term certificate: a certificate of $\Gamma_n$ certifies that $\Gamma_n$ is a node in covtree.)

\item A certificate of a node $\Gamma_n$ is also a certificate of every node below $\Gamma_n$.

\item Given a node $\Gamma_n$, repeated applications of $\mathcal{O}$ generate the unique path downwards from $\Gamma_n$ to the root.

\item In order to construct level $n$ of covtree, one considers all the non-empty subsets of $\Omega(n)$. These are the ``candidate nodes'' for level $n$. To determine whether a candidate node is a node in covtree one needs to determine whether it has a certificate. In general, this is a difficult problem.

\item Given any $n$-order $C_n$, the set $\{C_n\}$ is a node at level $n$ since $C_n$ is a certificate of $\{C_n\}$.

\item The first three levels of covtree are shown in Fig.\ref{covtree_fig_chap_2}. Levels 1 and 2 contain all candidate nodes, while level 3 contains 22 nodes out of 31 candidates. The $9$ ``non-nodes''  are shown in Fig.\ref{nonnodes}.
\end{itemize}

\subsection{The sample space, algebra and measure}\label{sample_subsec_chap2}

Our definition of \textit{certificate} implies that each $\lc{C}\in\lc{\Omega}$ is a labeled certificate of exactly one node at level $n$, for all $n>0$. The nodes of which $\lc{C}$ is a certificate form a path in covtree $\mathcal{P}=\Gamma_1\prec\Gamma_2\prec~\dots$, and this allows us to think of $\lc{C}$ as a certificate of the path itself:
\begin{definition}[Certificate of path]\label{cert_of_path_def}
An infinite order $C$ is a certificate of $\mathcal{P}$ if it is a certificate of every node in $\mathcal{P}$. A labeled certificate of $\mathcal{P}$ is a representative of a certificate of $\mathcal{P}$.
\end{definition}
We note that every $\lc{C}\in\lc{\Omega}$ is a labeled certificate of exactly one path. This ensures that every $\lc{C}\in\lc{\Omega}$ is contained in the covtree sample space, where the interpretation is that $\lc{C}$ is grown by the process when a random walker on covtree picks out the path of which $\lc{C}$ is a certificate. 

Consider a pair of order-isomorphic causal sets $\lc{C},\lc{D}\in\lc{\Omega}$. $\lc{C}$ and $\lc{D}$ are labeled certificates of the same nodes and are therefore associated with the same path. This is what we expected from a covriant dynamics: order-isomorphic causal sets (\textit{i.e.}, those causal sets which differ from each other only in their labels) cannot be distinguished. Therefore, instead of associating $\mathcal{P}$ with a growing causal set, one can associate $\mathcal{P}$ with a growing order $C=[\lc{C}]=[\lc{D}]$.

When $\lc{C}$ is a rogue, the class of causal sets associated with $\mathcal{P}$ is strictly larger than $[\lc{C}]$, since if $\lc{C}$ and $\lc{E}$ form a rogue pair then they are labeled certificates of the same nodes. This inability of covtree to distinguish between equivalent rogues suggests that the covariant $\sigma$-algebra on which a random walk on covtree defines a measure is the stem algebra, $\mathcal{R}(\mathcal{S})$, which was introduced in section \ref{sec_gd}. We shall see that this intuition is correct.

Before proceeding to consider the $\sigma$-algebra and measure in more detail, we must first satisfy ourselves that our interpretation of the nodes as sets of stems allows for every inextendible upward-going path $\mathcal{P}$ to be associated with some $\lc{C}\in\lc{\Omega}$. This is desirable for two reasons, interpretational and technical: it ensures that every realisation of the random walk can be associated with a growing causal set, and it guarantees that every set of transition probabilities on covtree yields a well-defined measure on the covariant event algebra.

 Let $\mathcal{P}$ denote an inextendible covtree path from the origin upwards, $\Gamma_1\prec\Gamma_2\prec \dots$

\begin{theorem}
Every path $\mathcal{P}$ has at least one certificate.
\label{theorem1}
\end{theorem}

\paragraph{Sketch of proof:} We will use the fact that for any $\Gamma_m\in\mathcal{P}$, there exists some $n > m$ such that $\Gamma_{n}\in\mathcal{P}$ contains some certificate $C_{n}$ of $\Gamma_{m}$ (see lemma 4.4 in \cite{Dowker:2019qiz}).

Choose any $\Gamma_l\in\mathcal{P}$ to begin with.

Pick some $\Gamma_{m}\in\mathcal{P}$ which contains a certificate $C_{m}$ of $\Gamma_{l}$. Pick a labeled representative $\lc{C}_{m}$ of ${C}_{m}$.

Pick some $\Gamma_{n}\in\mathcal{P}$ which contains a certificate $C_{n}$ of $\Gamma_{m}$. Pick a labeled representative $\lc{C}_{n}$ of ${C}_{n}$ in which $\lc{C}_{m}$ is a stem. This is always possible because $C_{m}$ is a stem in $C_{n}$.

Continue iteratively as above, at each stage picking a node which contains a certificate of the previous node and then picking a representative of this certificate with a labeling compatible with the previous labeled certificate.

This algorithm produces a countable sequence of labeled causets  $\lc{C}_{m}\subset  \lc{C}_{n}\subset~\dots $ whose union is a labeled certificate $\lc{C}$ of $\mathcal{P}$. The order $C$, of which $\lc{C}$ is a representative, is a certificate of $\mathcal{P}$. \hfill\(\Box\)

This establish the existence of a surjection from $\Omega$, the set of infinite orders, to the set of covtree paths. The upshot is that any realisation of a random walk on covtree can be identified with some history in $\Omega$---the growing order is a certificate of the path traced by the random walk.

As we already mentioned, a path will have more than one certificate if it is associated with rogue orders. How one should resolve this depends on the physical interpretation that one assigns to rogues. If one believes that rogues are unphysical and should never be grown by the process\footnote{Reasons to think this include that in the CSG models rogues never happen, \textit{i.e.} $\mu(\Theta)=0$ where $\Theta$ is the set of rogues, and that every rogue contains an infinite antichain corresponding to infinite space \cite{Brightwell:2002vw}.} then the resolution can be to only consider random walks in which the measure of the set of the ``rogue paths'' is null. An alternative is to allow rogues to arise but propose that they are physically indistinguishable (since they can only be distinguished globally, not by any local observer living on them). This is equivalent to replacing $\Omega$ with the space of rogue equivalence classes $\Omega/{\sim}_R$, where each path corresponds to exactly one class.

Another subtlety relates to the notion of growth. To what extent can we say that an order is \textit{growing} as the covtree walk advances? At stage $n$, we do not know which finite order has grown thus far nor its cardinality, only which $n$-stems it contains. While in the CSG models the growth is explicit, on covtree it is implicit or ``vague'' \cite{Wuthrich:2015vva}. But if there is a process of growth which can be associated with a covtree walk, then it may be that it is this quality of vagueness which embodies asynchronous becoming.

The surjection from $\Omega$ to the set of covtree paths does more than establish a narrative of growth. It enables us to use covtree to define a measure space of orders in the following way. To each covtree node assign its ``cylinder set''\footnote{``Cylinder set'' is a generic term in stochastic processes and should not be confused with its specific usage in \eqref{cyl_def_chap1}. The meaning should be clear from the context.}, the set of all paths $\mathcal{P}$ which contain it. The collection of all cylinder sets generates covtree's topological\footnote{It is called ``topological'' because the cylinder sets are the open balls under the metric topology given by the metric $d(\mathcal{P},\mathcal{P}')=1/2^n$, where $n$ is the number of nodes shared by $\mathcal{P}$ and $\mathcal{P}'$.} $\sigma$-algebra. Now, use the surjection which maps an infinite order to the path of which it is a certificate to pull back covtree's topological $\sigma$-algebra to a $\sigma$-algebra on $\Omega$. This pull-back algebra is the $\sigma$-algebra of observables in a covtree growth dynamics. The pull-back of the cylinder set associated with a given node $\Gamma_n$ is the set $cert(\Gamma_n)$, defined by,
\begin{definition}[Certificate set]\label{cert_set_def} For each covtree node $\Gamma_n$, its certificate set, $cert(\Gamma_n)$, is the set containing all its infinite certificates, \begin{equation}cert(\Gamma_n):=\{ C\in\Omega \mid C \text{ is a certificate of } \Gamma_n\}. \end{equation}\end{definition}
Thus, the $\sigma$-algebra of observables in the covtree growth dynamics is generated by the certificate sets. In our earlier discussion, we had already anticipated that this $\sigma$-algebra is $\mathcal{R}(\mathcal{S})$. Indeed, one can show that any stem set (cf. equation \eqref{stem_set_def}) can be constructed through a finite number of set operations on the certificate sets and vice versa. It follows that the collection of stem sets and the collection of certificate sets generate the same $\sigma$-algebra.

Finally, standard results in measure theory ensure that each covtree random walk (defined by a complete set of covtree transition probabilities) gives rise to a unique measure on $\mathcal{R}(\mathcal{S})$, where the measure of $cert(\Gamma_n)\in\mathcal{R}(\mathcal{S})$ is equal to the probability of reaching $\Gamma_n$ (\textit{i.e.} to the product of transition probabilities on the path from the root to $\Gamma_n$).

We had seen that any random walk on covtree gives a well-defined measure space of causal sets, and this completes our justification for interpreting covtree as a framework for growth dynamics.

We now have two methods for defining measures on $\mathcal{R}(\mathcal{S})$: via a restriction of a measure $\lc{\mu}$ on the labeled $\sigma$-algebra $\lc{\mathcal{R}}$, where $\lc{\mu}$ arises from a random walk on labeled poscau (shown in Fig.\ref{lposcau})\footnote{The random walk need not be a CSG model nor must the transition probabilities satisfy any physical conditions.}, or directly via a covtree random walk. It has been shown that the two methods give rise to the same class of measures, namely the class of measure on $\mathcal{R}(\mathcal{S})$. Every measure on $\mathcal{R}(\mathcal{S})$ can be derived from a covtree walk: the transition probability in the covtree walk from node $\Gamma_n$ to the node $\Gamma_{n+1}$ directly above it is the measure of $cert(\Gamma_{n+1})$ divided by the measure of $cert(\Gamma_{n+1})$. Additionally, every measure on $\mathcal{R}(\mathcal{S})$ possesses some (not necessarily unique) extension to $\tilde{\mathcal{R}}$ (see lemma 4.9 in \cite{Dowker:2019qiz}), meaning that every measure on $\mathcal{R}(\mathcal{S})$ can be obtained via a restriction of some $\lc{\mu}$. Thus, for every walk on labeled poscau---whether it satisfies discrete general covariance or not---there exists a covtree walk which produces the same measure on $\mathcal{R}(\mathcal{S})$. There is no easy relationship between the discrete general covariance condition on a labeled poscau walk and the manifest covariance of a covtree walk.

\section{The structure of covtree}\label{sec_furth_struc}
In section \ref{sample_subsec_chap2}, we had seen that a covtree walk is equivalent to a measure on $\mathcal{R}(\mathcal{S})$, and as such is a dynamics for causal sets. But there is no reason to expect that a generic covtree walk gives rise to a physically interesting dynamics: the class of covtree walks (or equivalently, the class of measures on $\mathcal{R}(\mathcal{S})$) is too vast to be interesting. We need physically motivated conditions to restrict the models to a sub-class worth studying.

The CSG models were derived by posing and solving two such conditions, and it is natural to consider how these conditions could be adapted to the covtree framework. However, when doing so, one comes across an obstacle: the formulation of the conditions satisfied by the CSG models relies on the use of labels to the extent that their potential generalisations to a label-free framework are obscured. This may be expected of the discrete general covariance condition, since its role---to impose invariance under relabeling---is redundant in a framework which makes no reference to labels. But as we saw in section \ref{sample_subsec_chap2}, the manifest covariance of covtree is not equivalent to the discrete general covariance condition, and what form discrete general covariance takes on the covtree transition probabilities is an interesting open question. The local causality condition satisfied by the CSG models (known as ``Bell causality'') states that the probability of transition from $\lc{C}_n$ to one of its children $\lc{C}_{n+1}$ depends only on the past the new-born element. The issue there is that one has to pin-point the new-born element, an impossible task when the objects considered are orders, not causal sets. As it stands, this tension between the global nature of label-independent objects and the local nature of causality is still in need of a resolve. We will return to it briefly in section \ref{sec_renorm}.

A complementary approach to identifying physical dynamics is requiring that the dynamics favour the physical kinematics (\textit{e.g.} requiring that manifold-like\footnote{We say an order $C$ is manifold-like if a representative of $C$ can be faithfully embedded into a four-dimensional Lorentzian manifold.} orders are likely to be grown). The success of translating such requirements into conditions on covtree transition probabilities hinges on understanding the relationship between paths and their certificates (\textit{e.g.} which paths have manifold-like certificates).

An understanding of the structure of covtree is also important for constraining the dynamics. For example, any dynamics should satisfy the Markov-sum-rule: the sum of the transition probabilities from any node $\Gamma_n$ must equal 1. But with no knowledge of the number of nodes directly above $\Gamma_n$ or of the relation they bear to it, this constraint is intractable. (In contrast, in the case of the CSG models knowing that the children of $\lc{C}_n$ in labeled poscau are in 1-to-1 correspondence with the stems in $\lc{C}_n$ allows to solve the Markov-sum-rule.)

In addressing these challenges, one might be tempted to construct covtree explicitly, but thus far only the first three levels of covtree have been worked out (Fig.\ref{covtree_fig_chap_2}). Brute force methods come up short in going to higher levels as the number of candidate nodes at level $n$ increases rapidly as $2^{|\Omega(n)|}-1$, where $|\Omega(3)|=5, |\Omega(5)|=63$ and $|\Omega(16)|=4483130665195087$ \cite{oeis}. However, progress has been made by focusing on structural properties which are independent of level. This section is dedicated to surveying these results. The interested reader may refer to \cite{Zalel:2020oyf} for their derivation.

\subsection{Nodes}
Here we list properties which pertain to nodes, including criteria for a set of $n$-orders to be a node, properties of minimal certificates and a study of direct descendants and valency. We begin with definitions.

\begin{definition}[Singleton and doublet] A node $\Gamma_n$ in covtree is a singleton if it contains a single $n$-order. A node $\Gamma_n$ in covtree is a doublet if it contains exactly two $n$-orders.\end{definition}

\begin{definition}[Covering causet/order]\label{covering_def} Given an $n$-causet $\lc{C}_n$, its covering causet $\widehat{\lc{C}_n}$ is the $(n+1)$-causet formed by putting the element $n$ above every element of $\lc{C}_n$. Similarly, $\widehat{{C}_n}$ is the covering order of $C_n$, where $\widehat{{\lc{C}}_n}$ and $\lc{C}_n$ are representatives of the respective orders. \end{definition}

\begin{figure}[htbp]
	\centering
	\includegraphics[width=0.4\textwidth]{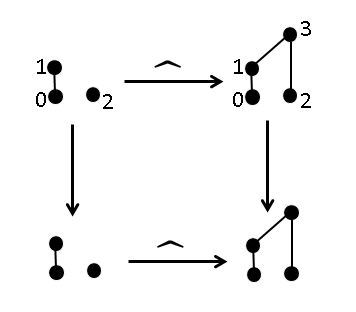}
	\caption[Covering causet/order]{The relationship between an order, its covering order and their representatives.}
	\label{cover}
\end{figure}

An illustration is shown in figure \ref{cover}.

Note that $C_n$ is the only $n$-stem in its covering order $\widehat{C_n}$,  and therefore the node $\{\widehat{C_n}\}$ is directly above $\{C_n\}$ in covtree. Thus,
\begin{property} \label{hat}
Every singleton has at least one direct descendant which is a singleton.
\end{property}

Moreover,
\begin{property} \label{unique_sing_above_sing_property}
If $\{\widehat{C_n}\}$ is the only singleton directly above $\{C_n\}$ then $\{\widehat{C_n}\}$ is the only node directly above $\{C_n\}$.\end{property}

Every singleton with valency greater than one has at least one direct descendant which is a doublet since:
\begin{property}\label{prop_22060201}
If $\{D_{n+1}\}\succ\{C_n\}$ and $D_{n+1}\not=\widehat{C_n}$ then $\{\widehat{C_n},D_{n+1}\}\succ \{C_n\}$.
\end{property}

A corollary of properties \ref{unique_sing_above_sing_property} and \ref{prop_22060201} is:
\begin{property}
No singleton has a valency of 2.
\end{property}

Singletons which possess property \ref{unique_sing_above_sing_property} are the only nodes in covtree which have exactly one direct descendant since: 
\begin{property} \label{valency}
Only singletons can have exactly one direct descendant in covtree.
\end{property}

Additionally, 
\begin{property}\label{prop_crown} For any $k\geq 1$ there is a singleton $\{C_n\}$ in covtree with $k$ singletons directly above it. \end{property}
An immediate corollary of property \ref{prop_crown}  is that the valency of singletons is unbounded. (Note that $k$ is not the valency of $\{C_n\}$, for if $k\geq 2$ then $\{C_n\}$ has additional direct descendants which are not singletons, cf. property \ref{prop_22060201}.)

\begin{figure}[htpb]
    \includegraphics[width=.8\textwidth]{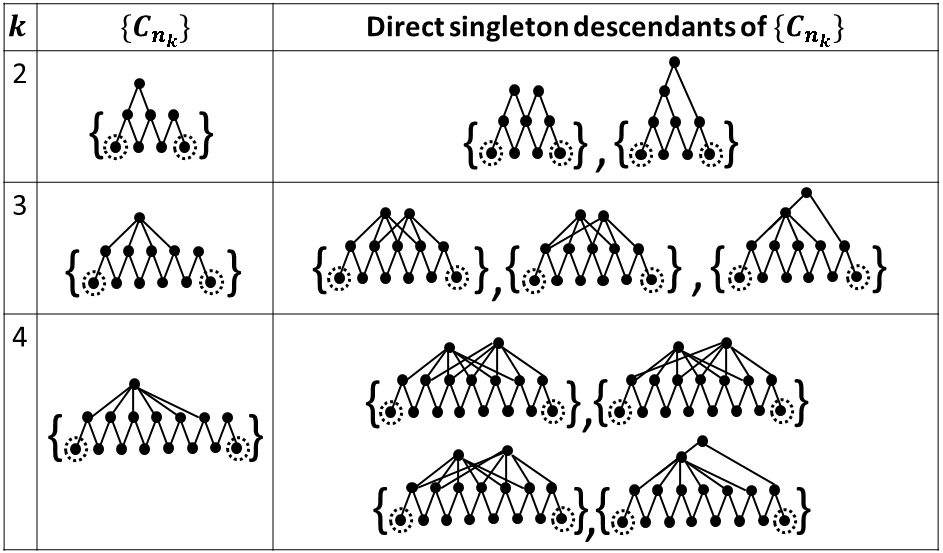}
  \caption[Crowns (property \ref{prop_crown})]{Illustration of property \ref{prop_crown}. The elements circled by a dotted line are identified with each other.}\label{singdesc_new}
\end{figure}

An example of a singleton node with 1 singleton directly above it is $\Gamma_4=\{$\twoch \twoch $\}$. To see that the statement is true for $k>1$, one can construct a countable sequence of singletons $$\{C_{n_2}\},\{C_{n_3}\},...,\{C_{n_k}\},...$$ such that $\{C_{n_k}\}$ has $k$ singletons directly above it. Fig.\ref{singdesc_new} shows the first three singletons in the sequence and their respective singleton descendants. 

Similarly,
\begin{property}\label{prop_200801} For any integer $k\geq 1$ there exists a doublet in covtree with $k$ singletons directly above it. \end{property}

\begin{figure}[htpb]
    \includegraphics[width=.9\textwidth]{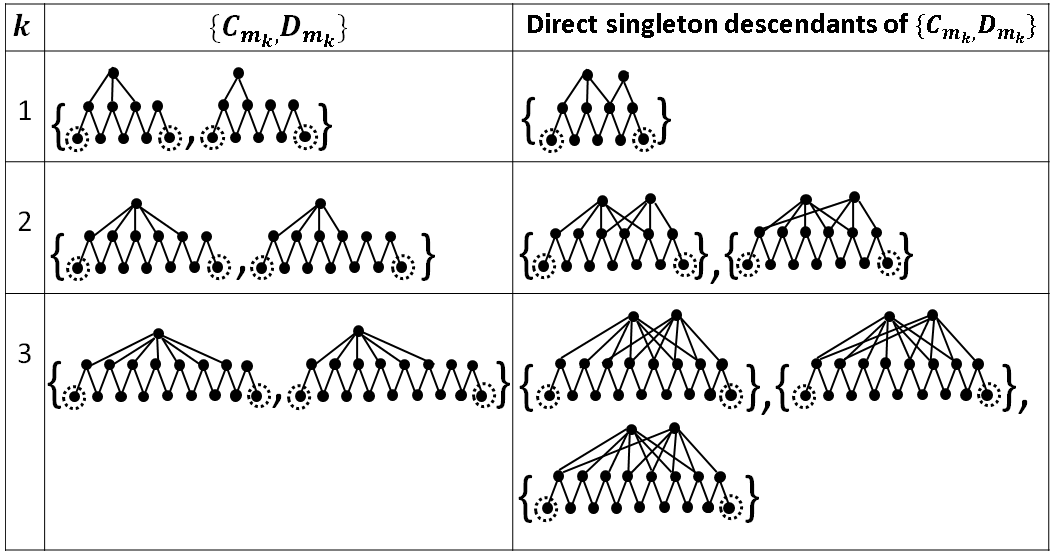}
   \caption[Crowns (property \ref{prop_200801})]{Illustration of property \ref{prop_200801}. The elements circled by a dotted line are identified with each other.}\label{fig_060904b}
\end{figure}

As before, one can construct an countable sequence of doublets, $$\{C_{m_1},D_{m_1}\},\{C_{m_2},D_{m_2}\},...,\{C_{m_k},D_{m_k}\},...$$ such that the $k^{th}$ doublet in the sequence has $k$ singletons directly above it. Fig.\ref{fig_060904b} shows the first three doublets in the sequence and their direct singleton descendants.


A key hurdle in the construction of covtree is understanding which sets of $n$-orders are covtree nodes. The following property gives a necessary condition in the case of doublets:
\begin{property}\label{prep_1307201}
$\{A_n,B_n\}$ is a doublet in covtree only if there exists an $(n-1)$-order $S$ which is a stem in both $A_n$ and $B_n$.
\end{property}

Property \ref{prop_270801} is a corollary:
\begin{property}\label{prop_270801}
If $\Gamma_n$ is a doublet in covtree then all minimal certificates of $\Gamma_n$ are $(n+1)$-orders.
\end{property}
Therefore, if $\Gamma_n$ is a doublet in covtree and $\Gamma_n\prec\Gamma_{n+1}$ then $\Gamma_{n+1}$ contains some minimal certificate of $\Gamma_n$. It is a corollary of properties \ref{prop_200801} and \ref{prop_270801} that for any integer $k\geq 1$ there exists a doublet in covtree with $k$ minimal certificates.

\subsection{Paths}\label{section_paths}
\begin{figure}[h]
  \centering
	\includegraphics[width=0.6\textwidth]{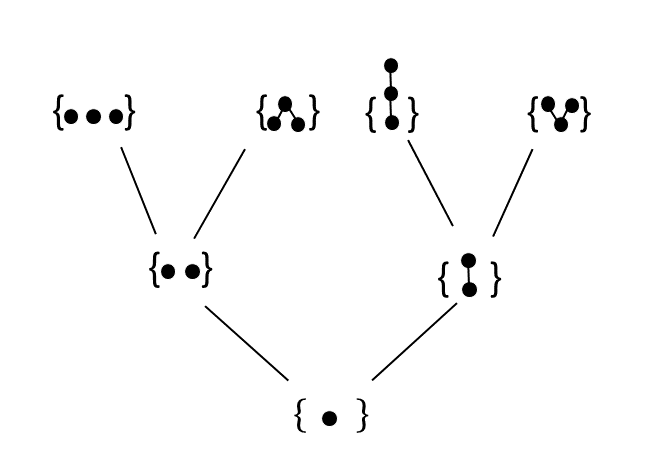}
	\caption[Singtree]{The first three levels of singtree.}
	\label{singtree}
\end{figure}

Here we present properties of certain covtree paths and their certificates.
\begin{property} In covtree, there are infinite upward-going paths from the origin in which every node is a singleton.\end{property}
We call the subset of covtree which contains exactly all these paths \textit{singtree}, since it is a tree of singletons. Fig.\ref{singtree} shows the first three levels of singtree.

To discuss singtree we will need the concept of the \textit{Newtonian order}.

\begin{definition}[Newtonian causet/order] A Newtonian causet is a causet in which every element in level $k$ is above every element in level $k-1$. A Newtonian order is an order whose representatives are Newtonian.\end{definition}

In a Newtonian causet, every pair of elements which are unrelated have the same past and the same future, alluding to a notion of a Newtonian global time, hence its name\footnote{Note however that a Newtonian order is \textit{not} a good approximation of continuum Euclidean space.}. A Newtonian causet is a ``stack of antichains'', and for any natural number $N$, the union of the first $N$ levels is a past of a break. The local finiteness condition implies that every level whose elements are not maximal must be finite.

One can show that an order $C$ is Newtonian if and only if for every natural number $n\leq |C|$ there is a unique $n$-order which is a stem in $C$. Thus, we have:

\begin{property}\label{prop_111022} A singleton $\{C_n\}$ is in singtree if and only if $C_n$ is Newtonian.\end{property}

\begin{property}\label{corollary_170901} An infinite order $C$ is Newtonian if and only if it is a certificate of a singtree path. \end{property}

If $\{C_n\}$ is a node in singtree then it has exactly two direct descendants in singtree: $\{\widehat{C_n}\}$ and $\{D_{n+1}\}$, where $D_{n+1}$ is the Newtonian order whose representative is constructed from a representative of $C_n$ by adding a new element to its maximal level. If $\{C_n\}$ is a node in singtree then it has exactly three direct descendants in covtree: its singtree descendants, $\{\widehat{C_n}\}$ and $\{D_{n+1}\}$, and the doublet $\{\widehat{C_n},D_{n+1}\}$.

Given property \ref{corollary_170901}, it is now a simple matter to solve for the family of covtree dynamics in which the set of non-Newtonian orders is null: it is the set of covtree walks in which the walker stays in singtree with probability 1, \textit{i.e.}
\begin{equation}
\mathbb{P}(\Gamma_n)=0 \ \forall \ \Gamma_n \text{ not in singtree.} 
\end{equation}
This family of Newtonian dynamics acts as a proof of principle, illustrating how an understanding of covtree could allow one to solve for a dynamics with particular features. But, since these dynamics are unphysical, this is very much a case of ``looking under the lamp-post''. Where are we to look if not under the lamp-post? One avenue for exploration is to ask: what role, if any, do rogues play in the physics of covtree walks?

Since in CSG models the set of rogues is null \cite{Brightwell:2002vw}, identifying covtree dynamics which possess this property is a step towards understanding what form CSG dynamics take on covtree. Moreover, if following \cite{Brightwell:2002vw} we are to choose $\mathcal{R}$ to be our $\sigma$-algebra of observables then---unless the covtree measure on $\mathcal{R}(\mathcal{S})$ has a unique extension to $\mathcal{R}$---one is faced with ambiguities both in interpretation and calculation. It is sufficient that the set of rogues be null for there to exist a unique extension, and therefore rogue-free dynamics are compatible with this approach.

One can draw an analogy between the condition that the set of rogues is null and the condition that the set of non-Newtonian orders is null: the former is the condition that the set of paths with more than one certificate is null, the latter the condition that the set of paths with more than one \textit{labeled} certificate is null. 
However, while we were able to solve for the latter, solving for the former poses a new challenge because it is a limiting condition: at no finite stage of the covtree walk can the claim that the growing order is a rogue be verfied or falsified. This is because for every node in covtree there exist both an infinite certificate which is a rogue and an infinite certificate which is not a rogue.

This means that there is no rogue analogue to singtree. Instead, we must look for other ways to obtain rogue-free dynamics. We will see in section \ref{sec_renorm} that pursuing the strictly stronger condition that the dynamics gives rise to infinitely many \textit{posts} or \textit{breaks} with unit probability is a promising route of particular interest for the causal set comology. 

\subsection{Self-similarity}\label{subsec_151202}
One of covtree's most interesting stuctural properties is its self-similarity. We now introduce this feature in advance of presenting its consequences for \textit{cosmic renormalisation} in section \ref{sec_renorm}.

Recall that covtree is itself a causal set whose ground-set is $\Lambda$ (definitions \ref{def_cert_sec_2} and \ref{covtree_def}).

\begin{definition}[Copy] A causal set $\Pi$ contains a copy of some causal set $\Phi$ if there exists a convex subcauset $\Phi'\subseteq \Pi$ such that $\Phi\cong\Phi'$. \end{definition}

\begin{definition}[Self-similar causal set] A causal set is self-similar if it contains infinitely many copies of itself. \end{definition}

For any finite order $A$, let $\Lambda_A\subset \Lambda$ be the convex subcauset of covtree which contains the node $\{\widehat{A}\}$ and everything above it.

\begin{theorem}\label{theorem_05082001}  For any finite order $A$, $\Lambda_A$ is a copy of covtree. Thus, covtree is self-similar. \end{theorem}

\begin{figure}[h]
	\centering
	\includegraphics[width=0.7\textwidth]{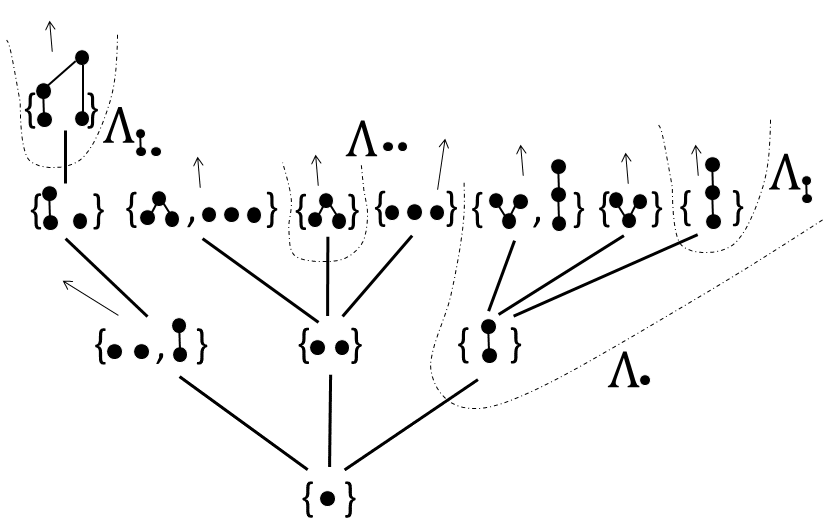}
	\caption[Covtree self-similarity]{The self-similar structure of covtree. The figure displays the first two levels of covtree in full and selected nodes from levels 3 and 4. The arrows indicate additional nodes not shown in the figure. The dashed lines indicate where a new copy of covtree begins. The ground-set $\Lambda_A$ of each copy is indicated next to each dashed line. Figuratively, we can write $\Lambda=\Lambda_{\emptyset}$.}
	\label{covtree_self_similarity}
\end{figure}

An illustration of covtree's self-similar structure is shown in Fig.\ref{covtree_self_similarity}.

The relationship between covtree and each of its copies $\Lambda_A$ is given by the map $\mathcal{G}_A$.
\begin{definition}[Break]\label{break_defn} A break in $\lc{C}$ is an ordered partition $\{\lc{A},\lc{B}\}$ of $\lc{C}$ such that $a\prec b \ \forall a\in\lc{A}, b\in\lc{B}$. $\lc{A}$ and $\lc{B}$ are called the past and future of the break, respectively. An order $C$ contains a break with past $A$ if a representative of it contains a break with past $\lc{A}$, where $\lc{A}$ is some representative of $A$.\end{definition}
 \begin{definition}[The map $\mathcal{G}_A$]\label{defn_161201} Given a finite order $A$ and a set $\Gamma_n\in\Lambda$, the map $\mathcal{G}_A$ takes $\Gamma_n$ to $\mathcal{G}_A(\Gamma_n)$, the set  of orders which contain a break with past $A$ and future $B_n\in\Gamma_n$, i.e.,
\vspace{-2mm}
\begin{equation*}
\mathcal{G}_A(\Gamma_n):=\{C \ | \ C \text{ is an order containing a break with past $A$ and future }B_n\in\Gamma_n\}.\end{equation*}\end{definition}

\begin{figure}[h]
	\centering
	\includegraphics[width=0.9\textwidth]{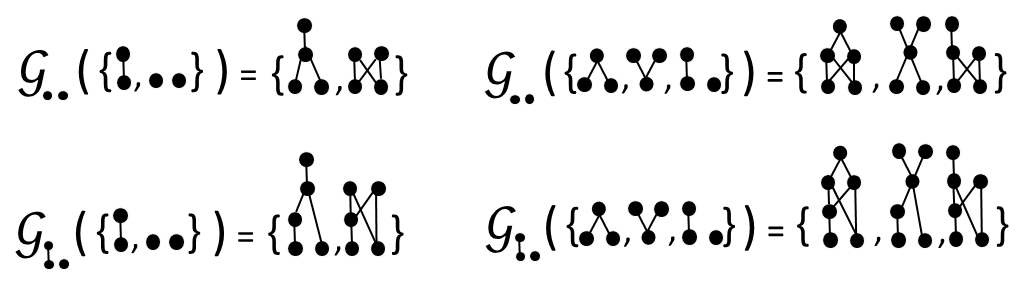}
	\caption[The map $\mathcal{G}_A$]{Illustration of the operation $\mathcal{G}_A$.}
	\label{G_A_defn}
\end{figure}

Examples are shown in Fig.\ref{G_A_defn}.

Covtree's self-similarity can be stated as: for any finite order $A$, the map $\mathcal{G}_A:\Lambda\rightarrow\Lambda_A$ is an order-isomorphism. The maps $\mathcal{G}_A$ are order-preserving because they commute with the map ${\mathcal O}$ (definition \ref{o_minus_defn}).

\section{Covtree and causal set cosmology}\label{sec_renorm}

An attractive lens through which to study growth dynamics is that afforded by the cosmological paradigm of \cite{Sorkin:1998hi} which aims to explain the emergence of a flat, homogeneous and isotropic cosmos directly from the quantum gravity era\footnote{This paradigm pertains only to the causal set spacetime, not to any matter living on it. Whether a causal set is enough to give rise to matter degrees of freedom \cite{Rideout:1999ub} or whether one requires additional structure such as a field living on the causal set is still unknown. Whichever the case may be, it is expected that this simplified cosmological paradigm will act as a guide to building a causal set cosmology.}. In this heuristic model, the fundamental parameters of nature change their values as the universe goes through subsequent epochs of expansion and collapse, echoing evolutionary mechanisms proposed by J. A. Wheeler, L. Smolin and others to explain the values of the parameters of nature \cite{Rees:1974,wheelerint,universeevolve,status}.

A causal set spacetime can be separated into epochs using the notions \textit{break} (definition \ref{break_defn}) and \textit{post}.

\begin{definition}[Post] $x\in\lc{C}$ is a post if it is related to all other elements in $\lc{C}$. The past (future) of a post  $x\in\lc{C}$ is its non-inclusive past (future) in $\lc{C}$. An order $C$ has a post with past $A$ if a representative of it contains a post with past $\lc{A}$, where $\lc{A}$ is some representative of $A$. \end{definition}

The notions of post and break are closely connected, since the following statements are equivalent:
$x$ is a post in $\lc{C}$; $\lc{C}$ admits a break $\{\lc{A},\lc{B}\}$ where $x$ is the unique maximal element of $\lc{A}$; and $\lc{C}$ admits a break $\{\lc{D},\lc{E}\}$ where $x$ is the unique minimal element of $\lc{E}$. 

In a growth dynamics, the parameters of nature are the transition probabilites themselves or a set of couplings from which they can be computed. The mechanism by which these parameters change their value from epoch to epoch is called \textit{cosmic renormalisation} \cite{Martin:2000js,Dowker:2017zqj}. The idea is that once the past of a break has been fully grown, the future of the break can be considered independently of the past as a growing causal set in its own right. The transition probabilities which govern the growth of the future are the ``renormalised'' probabilities---a repackaging of the original transition probabilities together with information about the partial order structure of the past.

When the renormalisation flow generated by successive epochs has certain features\footnote{For instance, that its stationary points grow causal sets with the desired cosmological features, that the basin of attraction of these stationary points is large and that it contains an abundance of dynamics which are likely to give rise to posts/breaks.}, the dynamics evolves into growing larger, flatter epochs as the universe cycles repeatedly. It is then only a matter of time until the universe displays the flat, homogeneous and isotropic feaures we observe today.

\subsection{Certificates with posts and breaks}
The cosmological narrative above gives us a broad class of dynamics to aim for: dynamics which are likely to grow causal sets with a large number of posts or breaks. Which covtree dynamics fall into this category is an open question, but a first step in answering it has already been achieved through the classification of covtree paths whose certificates contain posts or breaks.

Recall that $\widehat{A}$ denote the covering order of $A$ (definition \ref{covering_def}), and let $\Hat{\Hat{A}}$ denote the covering order of  $\widehat{A}$. 
\begin{theorem}\label{lemma131201}
Let the order $C$ be a certificate of the path $\mathcal{P}$. \begin{enumerate}\item $C$ contains a break with past $A$ if and only if $\{\widehat{A}\}$ is a node in $\mathcal{P}$. \item $C$ contains a post with past $A$ if and only if $\{\Hat{\Hat{A}}\}$ is a node in $\mathcal{P}$. \end{enumerate} \end{theorem}

With theorem \ref{lemma131201} in hand, the challenge ahead is to write down a complete set of covtree transition probabilities which are likely to lead the random walker through a long sequence of nodes of the form $\{\widehat{A}\}$. Returning to our discussion of rogue-free dynamics (cf. section \ref{section_paths}), note that a rogue causet contains an infinite level \cite{Brightwell:2002vw} and as a result cannot contain an infinite sequence of posts or breaks. Therefore, requiring that the random walker passes through infinitely many nodes of the form $\{\widehat{A}\}$ is not only cosmologically relevant, but also guarantees that the dynamics abhors rogue spacetimes.

\subsection{Cosmic renormalisation on covtree}

In addition to searching for dynamics which favour posts and breaks, constraints on the transition probabilities can be posed by requiring that the dynamics display certain behaviours under cosmic renormalisation. To do so, we must first outline the explicit form that cosmic renormalisation takes on covtree.

In the following, we denote a covtree transition probability by $\mathbb{P}(\Gamma_n\rightarrow\Gamma_{n+1})$. We use $\{\mathbb{P}\}$ to denote a complete set of covtree transition probabilities.

Consider a growing order $C$ which contains a break with past $A$. Theorem \ref{lemma131201} tells us that we can consider the past $A$ to have been fully grown when the random walker arrives at the node $\{\widehat{A}\}$. From this point onwards, we can consider the future as growing independently of this fixed past. We do so by acting on each node in $\Lambda_A$ with $\mathcal{G}_A^{-1}$ (the inverse of definition \ref{defn_161201}) since this effectively ``deletes'' the past $A$ of the break. $\mathcal{G}_A^{-1}$ maps $\Lambda_A$ to $\Lambda$ (cf. theorem \ref{theorem_05082001}) so that the growth of the future of the break (previously described by a walk on $\Lambda_A$) is now described as a walk on the whole of $\Lambda$ and is governed by a new set of \textit{effective} transition probabilities. Given a dynamics $\{\mathbb{P}\}$, the effective dynamics $\{\mathbb{P}_A\}$ which governs the growth of the future of a break with past $A$ is given by,
\begin{equation}\begin{split}\label{eq2507201}
R_A:\{\mathbb{P}\}\mapsto \{\mathbb{P}_A\}, \ \mathbb{P}_A(\Gamma_n\rightarrow\Gamma_{n+1})=\mathbb{P}(\mathcal{G}_A(\Gamma_n)\rightarrow\mathcal{G}_A(\Gamma_{n+1})).
\end{split}\end{equation}

Since the occurrence of a post with past $A$ is equivalent to the occurrence of a break with past $\widehat{A}$, the effective dynamics which governs the growth of the future of a post is given by transformation $R_{\widehat{A}}$, obtained from transformation $\eqref{eq2507201}$ via $A\rightarrow\widehat{A}$. An alternative formulation of renormalisation after a post can be obtained by considering the post to be the minimal element of the future of a break (rather than the maximal element of the past of a break).
In this case, the resulting effective dynamics is \textit{originary}, \textit{i.e.} $\mathbb{P}(\Gamma_1\rightarrow$\{\twoch\}$)=1$, reflecting the condition that all elements must be related to the post. We denote the associated transformation by $T_A$:\footnote{The apostrophe on the transition probabilities $\{\mathbb{P}'_A\}$ is used to distinguish between the images of $\{\mathbb{P}\}$ under $R_A$ and $T_A$.}
\begin{equation}\label{eq2407201}
\begin{split}
&T_A:\{\mathbb{P}\}\rightarrow\{\mathbb{P}'_A\},\\
&\mathbb{P}'_A(\Gamma_1\rightarrow\text{ \{\twoch\} })=1,\\
&\mathbb{P}'_A(\Gamma_n\rightarrow\Gamma_{n+1})=\mathbb{P}(\mathcal{G}_A(\Gamma_n)\rightarrow\mathcal{G}_A(\Gamma_{n+1})) \ \forall \  \Gamma_n \succeq \text{ \{\twoch\} },\\
&\mathbb{P}'_A(\Gamma_n\rightarrow \Gamma_{n+1})=0 \ \text{otherwise.}
\end{split}
\end{equation} 

\subsection{Cosmic renormalisation as a constraint}
The way in which the CSG models transform under cosmic renormalisation is well known \cite{Martin:2000js,Dowker:2017zqj}. In particular, it is known that the space of CSG models is closed under the cosmic renormalisation transformations, that there exists a unique one-parameter family of stationary points and that there are no higher order cycles. Additionally, the effective CSG dynamics depends on the past $A$ via two numbers only: the cardinality, $a$, and the number of maximal elements, $r$, of $A$. The remaining causal structure of $A$ is forgotten, and the various renormalisation transformations can be written in terms of powers of a single transformation, where the powers are simple functions of $r$ and $a$. This is reminicent of the form of the CSG transition probabilities which depend only on the cardinality and number of maximal elements of the past of the new-born element, a consequence of the Bell causality condition (cf. section \ref{sec_furth_struc}). 

Requiring that the cosmic renormalisation on covtree dynamics shares the features above could help us to better understand the form that the CSG models take on covtree. Additionally, new classes of physical dynamics could be obtained by requiring that covtree dynamics transform in particular ways. We summarise these ideas with some open questions: Is the condition on a covtree dynamics $\{\mathbb{P}\}$ that $\{\mathbb{P}_A\}=\{\mathbb{P}_B\}$ if and only if $A$ and $B$ have the same cardinality and number of maximal elements necessary for $\{\mathbb{P}\}$ to be a CSG dynamics? Is it sufficient? Does the factorisation property of the CSG transformations bear any relation to the constraint on a covtree dynamics $\{\mathbb{P}\}$ that, for any finite order $A$, the renormalisation transformation can be factorised as $R_A=R^{|A|}$ for some transformation $R$? When such a factorisation holds, the effective dynamics is independent of the causal structure of the past. Therefore, could the condition that $R_A$ factorises be interpreted as a causality condition on covtree dynamics?

\section{Variations}
The growth dynamics which we considered thus far, whether labeled or manifestly covariant, were constrained to grow those causal sets in which every element has a finite past. Thus, these dynamics can only describe cosmologies in which time has a beginning and it is natural to ask whether it is possible to construct growth dynamics for cosmologies in which time has no beginning \cite{Wuthrich:2015vva,Bento:2021voo}. From the outset, conceptual problems arise. Perhaps the most pressing of these is that, within the framework of labeled sequential growth, growing an infinite past requires that elements be born to the past of existing ones, making it (nearly if not entirely) impossible to conceive of the growth process as a physical phenomenon. However, one can identify a set of physically meaningful observables---namely the \textit{convex-events} which describe the convex suborders contained in the growing causal set---and this sets the stage for adapting covtree for two-way infinite growth. The variations presented here first appeared in \cite{Bento:2021voo}. 

\subsection{Terminology for two-way infinite causal sets}
A causal set is \textit{past-finite} (\textit{future-finite}) if every element is preceded (succeded) by at most finitely many others. A causal set is \textit{past-infinite} (\textit{future-infinite}) if it is not past-finite (future-finite). A causal set is \textit{two-way infinite} if it is both past-infinite and future-infinite.

In the previous sections, the sample space of our growth process was $\lc{\Omega}$, the set of labeled causal sets on the ground-set $\N$. Every infinite past-finite causal set is order-isomorphic to some $\lc{C}\in\lc{\Omega}$, or equivalently we can say that every countably infinite past-finite causal set has a natural labeling by $\N$. The converse is also true: only past-finite causal sets can have a natural labeling by $\N$.

To describe two-way infinite causal sets we must extend our index set from $\N$ to $\Z$. Every two-way infinite causal set has a labeling by $\Z$, though the converse is not true since past-finite causal sets with infinitely many minimal elements and future-finite causal sets with infinitely many maximal elements also admit labelings by $\Z$ \cite{honan:2018a,Gupta:2018}.

We generalise the definition of a labeled causet (definition \ref{def_labeled_causet}) to include those causal sets whose ground-set is an interval of integers (including the infinite intervals $\Z$ and $\N$) and whose partial order is compatible with the order on $\Z$ (\textit{i.e.} $x\prec y\implies x<y$). From here onwards, orders are defined to be order-equivalence classes of this extended class of labeled causal sets.

Let $C$ and $D$ denote orders with representatives $\lc{C}$ and $\lc{D}$, respectively. We will say that $C$ is a \textit{convex suborder} in $D$ if $\lc{D}$ contains a copy of $\lc{C}$. In that case we may also say that $C$ is a convex suborder in $\lc{D}$. If additionally $C$ is an $n$-order, we say that $C$ is an \textit{$n$-suborder} in $D$ or in $\lc{D}$. We will say that $C$ is a \textit{convex-rogue} if there exists another order $D\not\cong~C$ which has the same $n$-suborders as $C$ for all $n$. In that case we say that $C$ and $D$ are a convex-rogue pair. We may also refer to $\lc{C}$ and $\lc{D}$ as convex-rogues or as a convex-rogues pair.

\begin{figure}[h]
  \centering
	\includegraphics[width=0.3\textwidth]{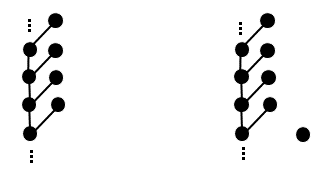}
	\caption[Convex-rogue pair]{The ``infinite comb'' (left) and the infinite comb disjoint union a single element (right) are convex-rogues since they contain the same convex suborders as each other.}
	\label{convex_rogues}
\end{figure}

An example of a convex-rogue pair is shown in Fig.\ref{convex_rogues}.

\subsection{Convex-covtree}
\label{sec:CovtreePastInfinite}

The first variation of covtree which we will encounter is \textit{convex-covtree}, whose definition is obtained from the definition of covtree by relacing \textit{$n$-stem} with \textit{$n$-suborder}. Thus, $\Gamma_n\subset\Omega(n)$ is a node in convex-covtree if and only if there exists some order $C$ whose set of $n$-suborders is $\Gamma_n$. We call $C$ the \textit{convex-certificate} of $\Gamma_n$. The ordering of the nodes in convex-covtree is as follows: for $m<n$, $\Gamma_m\prec \Gamma_n$ if $\Gamma_m$ is the set of $m$-suborders of the elements in $\Gamma_n$. One way to think about the ordering in convex-covtree is to pick an $n$-order in $\Gamma_n$ and delete a maximal or minimal element of it to form an $(n-1)$-order. Then $\Gamma_n$ is directly above the node $\Gamma_{n-1}$ that contains all $(n-1)$-orders which can be formed in this way.

\begin{figure}[htbp]
    \centering
    \begin{subfigure}{12cm} 
\includegraphics[scale=0.6]{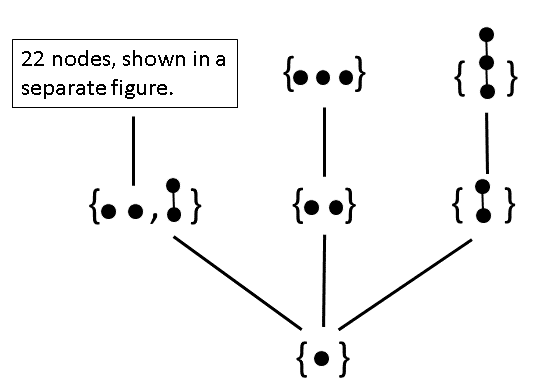}
    \caption{The first three levels of convex-covtree.}
    \end{subfigure}
    \begin{subfigure}{12cm}
     \includegraphics[scale=0.3]{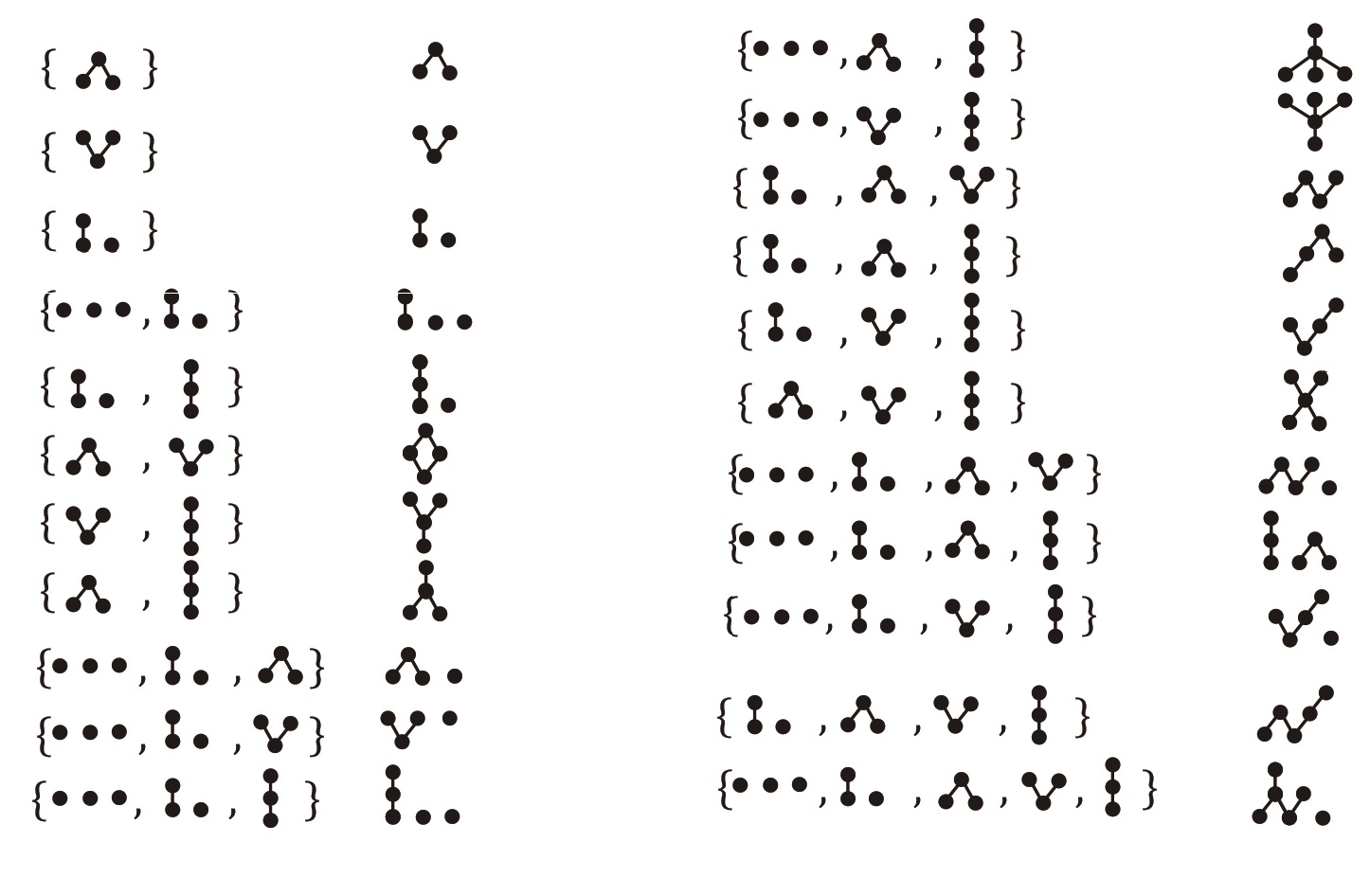}
    \caption{22 nodes of convex-covtree and their convex-certificates. These are the level 3 nodes which appear directly above the doublet.}
\end{subfigure}
            \caption[Convex-covtree]{The first three levels of convex-covtree.}
             \label{why}
\end{figure}

The nodes in the first three levels of convex-covtree are shown in Fig.\ref{why}.

Convex-covtree bears some similarities to covtree. In particular, every inextendible path in convex-covtree has a convex-certificate, allowing us to interpret a random walk on convex-covtree as a covariant process of growth: the growing order is a convex-certificate of the path which is traced by the random walk. Each node in the path corresponds to a covariant property of the growing order, \textit{i.e.} $\Gamma_n$ is the set of $n$-suborders of the growing order.

But unlike covtree, convex-covtree contains maximal nodes so that some of its inextendible paths are finite. This is a consequence of the fact that the existence of a finite-convex certificate does not guarantee the existence of an infinite one. In particular it is known that, if $C_n$ is not the $n$-chain or the $n$-antichain, the cardinality of the convex-certificates of $\{C_n\}$ is bounded from above by $n^2$. Finite inextendible paths are exactly the paths that contain such a singleton $\{C_n\}$. This does not mean that every such singleton is a maximal node, although the maximal nodes are always singletons containing their own---and their path's---unique convex-certificate. Thus, every finite inextendible path has a certificate. The converse is not true, some finite orders are certificates of no path at all.

Turning our attention to infinite paths, we note that every infinite order is a convex-certificate of some infinite inextendible path and conversely, every infinite path in convex-covtree has a convex-certificate (the proof is similar to that of theorem \ref{theorem1}). A path has more than one convex-certificate if its convex-certificates are convex-rogues and, in this case, which convex-certificate is the growing order is up for interpretation (\textit{e.g.} we can consider all convex-certificates of a given path to be physically equivalent).

\subsection{$\Z$-covtree}\label{subsec_inf_paths}
A second variation of covtree is $\Z$-covtree, defined as a trunctaion of convex-covtree: $\Z$-covtree is the subtree of convex-covtree which contains exactly all nodes that have a convex-certificate (with a representative) naturally labeled by $\Z$. Like covtree, $\Z$-covtree contains no maximal elements and every inextendible path has at least one convex-certificate. Moreover, one can show that every inextendible path has at least one convex-certificate whose representative is \textit{naturally labeled by $\Z$}, allowing us to consider the set of such orders as the sample space.

Like in covtree, the $\sigma$-algebra of observables is generated by the certificate sets associated with the nodes. For each $\Gamma_n$ in $\Z$-covtree, let $cert_{\Z}(\Gamma_n)$ denote the set of labeled convex-certificates of $\Gamma_n$ whose ground-set is $\Z$. A dynamics is given by a measure $\mu$ on the $\sigma$-algebra generated by the $cert_{\Z}(\Gamma_n)$'s, where $\mu(cert_{\Z}(\Gamma_n))=\mathbb{P}(\Gamma_n)$.

We saw in section \ref{sample_subsec_chap2} that covtree's observable algebra is equivalent to the stem algebra, $\mathcal{R}(\mathcal{S})$, generated by the stem sets of equation \eqref{stem_set_def}. We can pursue the analogy between stems and convex sets further by defining for each finite order $C_n$ the set $convex(C_n)$ to be the collection of labeled causets with ground-set $\Z$ which contain $C_n$ as a convex suborder. A \textit{convex-event} is any set which can be generated from the $convex(C_n)$'s via countable set operations (\textit{i.e.} a convex-event is an element of the $\sigma$-algebra generated by the $convex(C_n)$'s).  Each convex-event is a covariant measurable event with a clear physical meaning---it corresponds to a logical combination of statements about which finite orders are convex suborders in the growing causet. One can show that the $\sigma$-algebra generated by the $cert_{\Z}(\Gamma_n)$ is equal to the $\sigma$-algebra generated by the $convex(C_n)$.

The upshot is that $\Z$-covtree furnishes a growth framework for two-way infinite causal sets, with the caveat that  past-finite causal sets with infinitely many minimal elements and future-finite causal sets with infinitely many maximal elements must be suppressed by the dynamics.

\subsection{$\mathbb{N}$-covtree}\label{no_n_cov}
The success of $\Z$-covtree in providing a growth framework for two-way infinite causal sets based on the premise that the observables are convex-events raises the question: is it possible to define growth dynamics for past-finite causal sets in which the observables are convex-events? One can try doing so by defining a third variation of covtree, namely: the subtree of convex-covtree which contains exactly all nodes that have a convex-certificate (with a representative) naturally labeled by $\N$. We call this variation $\N$-covtree.

While the definition of $\N$-covtree is completely analogous to that of $\Z$-covtree, the resulting structure is not. In particular, there are inextendible paths $\mathcal{P}$ in $\N$-covtree which do not have a convex-certificate labeled by $\N$. By our definition of $\N$-covtree, every node in $\mathcal{P}$ has a convex-certificate labeled by $\N$---but there may be no such convex-certificate common to \textit{all} nodes in $\mathcal{P}$.

This means that convex-events cannot act as observables for past-finite causal sets since there is no surjection from the set of infinite past-finite orders to the set of $\N$-covtree paths and the measure space construction we described in section  \ref{sample_subsec_chap2} doesn't carry through.

One can understand this stark difference between $\Z$-covtree and $\N$-covtree using the language of metric spaces. For any two orders $C$ and $D$, let $C\sim D$ if and only if $C$ and $D$ are a convex-rogue pair, \textit{i.e.} if they share the same $n$-suborders for all $n$. Let $\Omega_{\mathbb{N}}$ and $\Omega_{\mathbb{Z}}$ denote the sets of orders which have a representative with ground-set $\N$ and $\Z$, respectively. Let $\Omega_{\mathbb{N}}/\sim$ and $\Omega_{\mathbb{Z}}/\sim$ be quotient spaces under the convex-rogue equivalence relation, so that their elements are equivalence classes of orders denoted by $[C]$ \textit{etc.} We can consider these quotient spaces as metric spaces with metric $d([C],[D])=\frac{1}{2^n}$, where $n$ is the largest integer for which representatives of $[C]$ and $[D]$ have the same sets of $n$-suborders. Given a node $\Gamma_n$ in convex-covtree we can associate with it a subset $[cert_{\mathbb{N}}(\Gamma_n)]\subseteq\Omega_{\mathbb{N}}/\sim$, namely the set of elements of $\Omega_{\mathbb{N}}/\sim$ whose representatives are convex-certificates of $\Gamma_n$, and similiarly $[cert_{\mathbb{Z}}(\Gamma_n)]\subseteq\Omega_{\mathbb{Z}}/\sim$. Given a convex-covtree path $\mathcal{P}=\Gamma_1\prec\Gamma_2\prec...$, we can associate with it the sets $[cert_{\mathbb{N}}(\mathcal{P})]=\bigcap_{\Gamma_n\in\mathcal{P}}[cert_{\mathbb{N}}(\Gamma_n)]$ and $[cert_{\mathbb{Z}}(\mathcal{P})]=\bigcap_{\Gamma_n\in\mathcal{P}}[cert_{\mathbb{Z}}(\Gamma_n)]$. The metric space $(\Omega_{\mathbb{Z}}/\sim,d)$ is complete, and therefore by Cantor's lemma $[cert_{\mathbb{Z}}(\mathcal{P})]$ is non-empty whenever all the $[cert_{\mathbb{Z}}(\Gamma_n)]$ is non-empty for all $\Gamma_n\in\mathcal{P}$. On the other hand, the metric space $(\Omega_{\mathbb{N}}/\sim,d)$ is not complete and therefore $[cert_{\mathbb{N}}(\mathcal{P})]$ can be empty even when $[cert_{\mathbb{N}}(\Gamma_n)]$ is non-empty for all $\Gamma_n\in\mathcal{P}$.

\begin{figure}[htpb]
    \centering
    \includegraphics[scale=0.34]{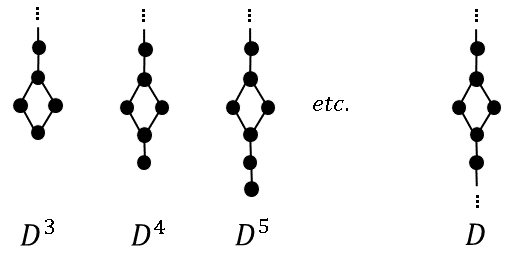}
    \caption[The non-existence of $\mathbb{N}$-covtree]{The order $D\in\Omega_{\Z}$ shown on the right is a convex-certificate of the path $\mathcal{P}$. Every node in $\mathcal{P}$ has a convex-certificate in $\Omega_{\mathbb{N}}$: $D^3$ is a convex-certificate of $\Gamma_n\in\mathcal{P}$ only for $n\leq 3$, $D^4$ is a convex-certificate of $\Gamma_n\in\mathcal{P}$ only for $n\leq 4$, $D^5$ is a convex-certificate of $\Gamma_n\in\mathcal{P}$  only for $n\leq 5$, \textit{etc}. There is no order in $\Omega_{\mathbb{N}}$ which is a convex-certificate of every node in $\mathcal{P}$.}
    \label{path_cert_fig}
\end{figure}

For example, consider the path,\begin{equation}\label{eq_path_ex}
\mathcal{P}=\{\ \oneach \}\prec \{\twoch, \ \twoach\}\prec\{\threech, \lambdacauset, \vee  \}\prec\{ \fourch, \lambdafour,\topvee , \diamond \ \}\prec ... 
\end{equation}
Each node $\Gamma_n\in\mathcal{P}$ has a convex-certificate $D^n\in\Omega_{\mathbb{N}}$, as illustrated in Fig.\ref{path_cert_fig}. These convex-certificates (technically, the equivalence classes in $\Omega_{\mathbb{N}}/\sim$ of which they are representatives) form a Cauchy sequence in $(\Omega_{\mathbb{N}}/\sim,d)$, where $d([D^n],[D^{n-1}])=\frac{1}{2^n}$. The limit of the sequence is the order $D$ shown in Fig.\ref{path_cert_fig}. Since $D$ is two-way infinite, we know that $[D]\not\in\Omega_{\mathbb{N}}/\sim$ so that  $(\Omega_{\mathbb{N}}/\sim,d)$ is not a complete metric space. Additionally, $D$ is the only certificate of $\mathcal{P}$, so $\mathcal{P}$ is an example of a path in $\N$-covtree which has no past-finite convex-certificate.

Finally, note that the machinery of metric spaces can be used to give an alternative proof to theorem \ref{theorem1} which stated that every inextendible path in covtree has a certificate. In this case, the metric space is $(\Omega_{\mathbb{N}}/\sim_R,\delta)$ where $\sim_R$ is the rogue equivalence relation (cf. definition \ref{rog_def}) and $\delta([C]_R,[D]_R)=\frac{1}{2^n}$ where where $n$ is the largest integer for which representatives of $[C]_R$ and $[D]_R$ have the same sets of $n$-stems.

\section{Discussion}\label{conc_chap}
Covtree and its variations form a manifestly covariant, label-independent framework through which growth dynamics for causal sets can be defined. Their study is motivated by the need to understand general covariance within quantum gravity, and one approach to doing so is to ask whether one can formulate the laws of physics in a way which makes reference only to physical (and not to gauge) degrees of freedom. Modern theoretical physics has thus far favoured gauge theories, but covtree is proof that at least within the discrete setting of causal set theory it is possible to do away with gauge degrees of freedom. In future, the unified nature of quantum gravity may also offer new possibilities in this direction. A second motivation for the development of these covariant dynamics has been that the labeled sequential growth dynamics have thus far resisted quantization and there is hope that a covariant formulation may offer a new route to quantum dynamics. Indeed, a label-independent formulation may prove necessary since concepts unrelated to each other in our current theories, such as general covariance and quantum interference, may prove inseparable in a full theory of quantum gravity.

One of the interesting issues which are highlighted by covtree is the interplay between the notions of ``local'' and ``global''. One can consider a causal set as a local object and an order as its global counterpart, since in a causal set one can identify individual elements and in an order one cannot. Similarly, in the labeled sequential growth it is known exactly which element is born at each stage of the growth but in a covtree growth such an element cannot be identified in general. On the other hand, there is also a certain flavour of locality in covtree since at each finite stage of the covtree process we know which stems are contained in the growing causal set but we don't know how they fit together (there is no God's eye view, only the viewpoint of somewhat local observers). In a similar vein, one can ask whether the failure of the convex-events to form a set of observables for past-finite causal sets can be interpreted as a statement about the local/global nature of observables: the event that the growing causal set contains some $n$-suborder pertains to the whole of the causal set, but the statement that it contains some $n$-stem is anchored to the antichain of minimal elements.

The condition that the causal set contains a break or a post is a global condition, since it pertains to every element in the causal set. As a result, the occurrence of a break or a post with a given past can be falsified but never verified at a finite stage of the (inherently local) sequential growth dynamics, forcing our hand to perform post-selection in order to discuss cosmic renormalisation. This post-selection is no longer necessary in the covtree framework, since the occurrence of a break or a post with a given past is synonymous with the random walker passing through a particular node of the form $\{\widehat{A}\}$. We find that the global occurrence of a break or a post manages to give us a glimpse of locality in the covtree process since it is exactly when the walker passes through one of these nodes that one can discern which causal set has been grown thus far---$\widehat{A}$---and identify a new born element---the maximal element of $\widehat{A}$. It is then that one can most convincingly associate a notion of growth with the covtree random walk.

This forms yet another motivation---in addition to those provided by the causal set cosmological paradigm and by the search rogue-free dynamics---to seek covtree dynamics which give rise to an infinite sequence of breaks or posts.

\section{Cross-References}
Discrete Dynamics of Sequential Growth, by Rideout, D (Chapter ID: 74)

\bibliographystyle{unsrt}
\bibliography{chapter_bib}{}

\end{document}